\documentclass[aps,prb,twocolumn,showpacs,amsmath]{revtex4}
\usepackage{latexsym}
\usepackage{amssymb}
\usepackage{graphicx}
\usepackage{bm}
\usepackage[english]{babel}
\usepackage[latin1]{inputenc}
\usepackage{amsmath}
\usepackage{amsfonts}

\newcommand{\pdag}{{\phantom{\dagger}}}
\newcommand{\bq}{\begin{equation}}
\newcommand{\eq}{\end{equation}}
\newcommand{\bn}{\begin{eqnarray}}
\newcommand{\en}{\end{eqnarray}}

\begin{document}

\title{Time-dependent quantum transport through an interacting quantum dot beyond sequential tunneling: second-order quantum rate 
equations}

\author{Bing Dong}
\affiliation{Key Laboratory of Artificial Structures and Quantum Control (Ministry of Education), Department of Physics, Shanghai 
Jiaotong University, 800 Dongchuan Road, Shanghai 200240, China}

\author{G. H. Ding}
\affiliation{Key Laboratory of Artificial Structures and Quantum Control (Ministry of Education), Department of Physics, Shanghai 
Jiaotong University, 800 Dongchuan Road, Shanghai 200240, China}

\author{X. L. Lei}
\affiliation{Key Laboratory of Artificial Structures and Quantum Control (Ministry of Education), Department of Physics, Shanghai 
Jiaotong University, 800 Dongchuan Road, Shanghai 200240, China}

\begin{abstract}

A general theoretical formulation for the effect of a strong on-site Coulomb interaction on the time-dependent electron transport 
through a quantum dot under the influence of arbitrary time-varying bias voltages and/or external fields is presented, based on 
slave bosons and the Keldysh nonequilibrium Green's function (GF) techniques. To avoid the difficulties of computing double-time 
GFs, we generalize the propagation scheme recently developed by Croy and Saalmann to combine the auxiliary-mode expansion with 
the celebrated Lacroix's decoupling approximation in dealing with the second-order correlated GFs and then establish a closed set 
of coupled equations of motion, called second-order quantum rate equations (SOQREs), for exact description of transient dynamics 
of electron correlated tunneling. We verify that the stationary solution of our SOQREs is able to correctly describe the Kondo 
effect on a qualitative level. Moreover, a comparison with other methods, such as the second-order von Neumann approach and 
Hubbard-I approximation, is performed. As illustrations, we investigate the transient current behaviors in response to a step 
voltage pulse and a harmonic driving voltage, and linear admittance as well, in the cotunneling regime.

\end{abstract}

\date{\today}


\maketitle

\section{Introduction}

The dynamical response of conduction electrons in open nanostructures driven by time-dependent external fields has attracted 
tremendous interest in coherent quantum transport theory because of its potential application to quantum information processing 
and single-electron 
devices.\cite{Fujisawa,Pekola,Gabelli,Feve,Switkes,DiCarlo,Watson,Leek,Wiel,Kouwenhoven,Blumenthal,Nevou,Hohls} In experiments, 
one is interested in measuring the current response of a mesoscopic system, usually a quantum dot (QD), driven by time-dependent 
periodic signals applied either on the system itself or on the attached leads, leading to the observation of the phenomena, such 
as quantum pumping\cite{Switkes,DiCarlo,Watson,Leek} and photon-assisted tunneling.\cite{Wiel,Kohler} An alternative way of 
external field modulation is to apply radio-frequency signals to the metallic gates that control the opening of the QD to its 
electrodes by oscillating tunneling barriers, i.e. a quantum dot turnstile.\cite{Kouwenhoven,Blumenthal,Nevou,Hohls} It is known 
that the fundamental physics of the nanoscale structures is dominated by two effects, the quantum coherence and electron 
correlations, which significantly affect the electronic tunneling through the nanostructures, giving rise to prototypical quantum 
phenomena, for instance, the Coulomb blockade and the Kondo effect.
These time-resolved experiments have therefore motivated the theoretical investigations for transient current-response behavior 
of an interacting QD under influences of arbitrary time-varying external fields in the cotunneling and the Kondo regimes.

In literature, various theoretical methods have been developed to study these time-dependent nonequilibrium phenomena. The 
scattering matrix approaches can only be applicable for the noninteracting system.\cite{Brouwer,Moskalets} The nonequilibrium 
Green's function (NGF) methods are believed to be the most powerful method for investigation of nonequilibrium time-dependent 
transport problem in an arbitrary nanoscale 
device.\cite{Jauho,Stefanucci,Zhu,Maciejko,Zheng,Moldoveanu1,Xing,Schiller1,Lopez1,Lopez2,Goldin,Kaminski,Nordlander,Plihal,Schil
ler2,Schmidt} For the case of noninteracting system, exact analytical solutions were obtained based on the time-dependent 
nonequilibrium Green's function (TDNGF) to the nonlinear transient current through a QD in response to sharp step- and 
square-shaped voltage pulses and harmonic bias modulations as well.\cite{Jauho,Maciejko} For an interacting system, certain 
approximations have to be invoked to calculate the correlated GFs. For instance, the conventional perturbation theory in the 
on-site interaction strength $U$ was developed to study the transient dynamics of the Anderson impurity 
model.\cite{Lopez2,Schmidt} Ac response of this model in the Kondo regime was examined by applying the time-dependent 
Schrieffer-Wolff transformation and combining with the perturbation theory,\cite{Schiller1,Goldin,Kaminski} or using an 
interpolation of the retarded self-energy.\cite{Lopez1} An exact analytical solution for a Kondo QD in response to a rectangular 
pulsed bias potential was also derived in the Toulouse point.\cite{Schiller2} Departure from this special point, numerical 
investigations on this issue were performed using time-dependent noncrossing approximation (NCA).\cite{Nordlander,Plihal}
In recent years, several advanced approaches that can deal with higher order dynamics have been developed based on the real-time 
quantum Monte Carlo and iterative path-integral methods,\cite{Muhlbacher,Weiss,Segal}, real-time density matrix renormalization 
group methods,\cite{Heidrich,Boulat,Guttge} etc. Albeit that these new methods are favorable in studying strong correlation 
effects, the validity of their application in the time-dependent nonequilibrium problem is still controversial.

The generalized quantum master equations (GQMEs) for the reduced density 
matrix,\cite{Schoeller,Splettstoesser,Cavaliere,Gurvitz,Dong1,Dong2,Dong3,Pedersen,Harbola,Welack,Timm,Jin,Li,Moldoveanu2,Koller} 
have the ability to cover the whole parameter regions of the interaction strength, from weak Coulomb interactions (sequential 
tunneling in the Coulomb blockade regime),\cite{Gurvitz,Dong1,Dong2,Harbola,Welack} intermediate couplings 
(cotunneling),\cite{Schoeller,Dong3,Pedersen,Timm,Jin,Koller} to strong Coulomb interactions (the Kondo-type 
tunneling),\cite{Schoeller,Li} depending on which order in perturbation expansion with respect to the tunneling Hamiltonian is 
evaluated in the transport kernels. The real-time diagrammatic technique provides a systematic and unified scheme to handle any 
order of expansion terms in tunneling Hamiltonian and can be in principle exact for quantum transport.\cite{Schoeller} This 
technique was then applied to study adiabatic and nonadiabatic pumping problems in a QD.\cite{Splettstoesser,Cavaliere} It is 
nevertheless a formidable task to identify all diagrams involving higher order contributions in this formalism. Recently, an 
evaluation of the transport kernels up to fourth-order expansion was accomplished for the stationary tunneling, which can account 
for all processes involving coherent tunneling of one or two electrons.\cite{Koller} Actually, certain kind of contributions 
beyond fourth-order transport kernels can be included by expanding equations of motion for the density matrix and the first-order 
auxiliary current matrices.\cite{Pedersen} This second-order von Neumann (SOvN) approach however relies on the Markov limit, and 
then fails to describe the transient tunneling dynamics. A similar but more rigorous scheme was developed by the hierarchical 
equations of motion (HEOM) based on the influence functional in the path integral formalism.\cite{Jin} For infinite levels the 
HEOM is exact and its second-tier approximation has been proven to be completely in agreement with the results by the real-time 
diagrammatic formalism at the same level.\cite{Jin} Recently, the HEOM has been utilized to study the Kondo effect in the 
stationary tunneling.\cite{Li} Therefore, a well developed and convenient theoretical formalism deserves to be further improved 
for the description of transient dynamics of strongly correlated nanoscale structures under arbitrary time-dependent external 
fields, and is currently a very active research area.

More recently, a general propagation scheme was developed by Croy and Saalmann for the transient problem in a noninteracting QD 
based on the TDNGF techniques.\cite{Croy1} In a similar fashion as the SOvN, they started from the equations of motion (EOMs) of 
the density matrix and auxiliary current matrices. By virtue of the TDNGF techniques and an auxiliary-mode expansion, the current 
matrices can then be expressed in a closed form. To this end, a set of coupled differential equations for quantities with only 
one time argument was obtained. In this way, this scheme avoids huge costs of numerical calculations in the traditional 
approaches, which compute directly the double-time GFs in time-domain requiring a double integration over 
time.\cite{Zhu,Moldoveanu1} Meanwhile, this scheme allows us to study arbitrary time dependences and structured reservoirs. This 
renders it an ideal tool for the time-dependent problem. Then this scheme was applied to study electron pumps\cite{Croy4} and 
time-dependent transport in molecular junctions.\cite{Zheng2,Xie,Popescu} Moreover, this scheme was extended to the interacting 
case in the Hubbard-I approximation.\cite{Croy2}
The purpose of this paper is therefor to generalize this propagation scheme to account for the electron-electron interaction 
beyond this level of approximation, making it capable of investigating the strong correlation effects in time-dependent 
nonequilibrium situations, e.g. the cotunneling and dynamic Kondo-type tunneling.

The remainder of this paper is organized as follows. In Sec. II we introduce the time-dependent Anderson model Hamiltonian and 
its slave particle representation. In Sec. III we present the theoretical methodology. Starting from the EOM of density matrix, 
we define the auxiliary current matrices $\Pi_{\eta\sigma}(t)$ and then expand them in terms of auxiliary modes by employing a 
decomposition of the Fermi function and NGF technique in the wide band limit. Subsequently, we derive the EOM of the 
auxiliary-mode expansion of the current matrices, which will generate the next-order auxiliary density matrices (AMDEs) being 
related to the second-order correlated GFs due to the on-site Coulomb interaction. To truncate this procedure, a decoupling 
approximation has to be invoked. In this paper, we will utilize three decoupling approximations to calculate the second-order 
GFs, including the Hubbard-I approximation, the Lacroix's decoupling approximation,\cite{Lacroix} and its simplified 
version.\cite{Meir} We find that the EOM of the current matrices is already closed at the Hubbard-I approximation and is 
consistent with the previous result.\cite{Croy2}. While within the Lacroix's decoupling approximation, the whole procedure will 
be closed in the EOM of the second-order AMDEs. That is the reason that we call them as the second-order quantum rate equations 
(SOQREs). In Sec. IV, in order to appreciate our formulism, we use our resulting SOQREs to study the steady-state transport 
through an interacting QD in the Kondo regime and compare our results with the previous exact solutions of EOM of retarded GFs at 
the same level of approximation.\cite{Entin,Kashcheyevs,Roermund} In Sec. V, we apply our SOQREs to examine transient current in 
response to steplike and harmonic modulations of bias voltages, and ac linear response as well, in cotunneling regime. Finally, a 
conclusion is presented in Sec. VI.

\section{Model Hamiltonian and slave-particle representation}

We consider an interacting quantum dot with a single energy level connected to two normal metal leads, whose Hamiltonian can be 
separated as three parts:
\bq
H=H_{C}+H_{QD} + H_T. \label{ham}
\eq
The Hamiltonian $H_C$ describes the noninteracting electrons in the two reservoirs, the left (L) and right (R) leads, and takes 
the following form:
\bq
H_C=\sum_{\eta k \sigma} \epsilon _{\eta k}(t) c_{\eta k\sigma }^{\dagger }c_{\eta k\sigma }^{\pdag},
\eq
where $\eta=\{{\rm L,R}\}$ stands for lead index, $k$ denotes electron momentum, and $\sigma=\{\uparrow,\downarrow\}$ is the spin 
orientation of electrons. Here $c_{\eta k \sigma}^{\dagger}$ and $c_{\eta k \sigma }$ are electron creation and annihilation 
operators for the electronic state $k \sigma$ in the lead $\eta$, respectively. Each of the two leads is separately in thermal 
equilibrium with the chemical potential $\mu_{\eta}$, which is set to be zero in equilibrium condition and chosen as the energy 
reference throughout the paper.
The QD Hamiltonian adopts the single-impurity Anderson model,
\bq
H_{QD} =  \sum_{\sigma} \epsilon_{d\sigma}(t) c_{d \sigma }^{\dagger }c_{d \sigma }^{\pdag} + Un_{d \uparrow }n_{d \downarrow }, 
\label{qd}
\eq
in which $c_{d \sigma}^{\dagger}$ and $c_{d \sigma}$ are creation and annihilation operators for a spin-$\sigma$ electron with an 
energy $\epsilon_{d\sigma}(t)$ on the QD. $n_{d\sigma}=c_{d \sigma}^{\dagger} c_{d \sigma}^{\pdag}$ is the occupation operator in 
the QD and $U$ is the charge energy. In this paper, we assume that the charge energy is large enough to exclude the double 
occupancy of the dot level. The last term $H_T$ describes the tunnel coupling between the QD and two electrodes,
\bq
H_T =\sum_{\eta k \sigma} \left ( V_{\eta k \sigma}(t) c_{\eta k\sigma }^{\dagger }c_{d \sigma}^{\pdag} +{\rm {H.c.}} \right ), 
\label{tunneling}
\eq
with the coupling matrix element $V_{\eta k \sigma}(t)$. Here to model time-dependent transport problem subject to arbitrary 
time-varying bias and/or gate voltages, the single-particle energies in the dot $\epsilon_{d\sigma}(t)$, in the leads 
$\epsilon_{\eta k}(t)$, and the coupling matrix element $V_{\eta k \sigma}(t)$ are introduced to be all time dependent and 
independent of each other: $\epsilon_{d\sigma}(t)=\epsilon_{0\sigma}+ \Delta_{d}(t)$, $\epsilon_{\eta k}(t)=\epsilon_{\eta k}^0+ 
\Delta_{\eta}(t)$, and $V_{\eta k \sigma}(t)=u_{\eta}(t) V_{\eta k \sigma}^0$. Without loss of generality, we assume that 
$\Delta_{d}(t)$, $\Delta_{\eta}(t)$, and $u_{\eta}(t)$ are all real functions of time.
Throughout we will use natural units $e=\hbar=k_{\rm B}=1$.

According to the infinite-$U$ slave-boson representation,\cite{Zou} the electron operator $c_{d \sigma}$ can be written in three 
possible single electron states, namely: the empty state $|0\rangle$ with zero energy $\varepsilon_{0}=0$, the singly occupied 
(with spin up or down) electronic state $|\sigma \rangle$, as
\bq
c_{d \sigma}=|0\rangle \langle \sigma | . \label{sf}
\eq
Because these three states expand the entire Hilbert space, the completeness relation must be satisfied
\bq
|0\rangle \langle 0|+ \sum_{\sigma} |\sigma \rangle \langle \sigma |=1. \label{comp}
\eq
These Dirac brackets were then treated as operators: $b^{\dagger}=|0\rangle$ as slave-boson operator and 
$f_{\sigma}^{\dagger}=|\sigma \rangle$ as pseudo-fermion operator. In terms of these auxiliary operators, Eqs.(\ref{sf}) and 
(\ref{comp}) become
\bn
&& c_{d\sigma}=b^{\dagger} f_{\sigma}^{\pdag} , \label{sf2} \\
&& b^\dagger b + \sum_{\sigma}f_\sigma^\dagger f_\sigma =1.
\label{comp2}
\en
The explicit (anti)communicators of these auxiliary particles can be easily established from the definitions of the Dirac 
brackets:\cite{Guillou}
\bq
bb^{\dagger}=1, \, f_{\sigma}^{\pdag} f_{\sigma'}^{\dagger}=\delta_{\sigma \sigma'}, \, 
bf_{\sigma}^{\dagger}=f_{\sigma}^{\pdag}b^{\dagger} = 0. \label{quan}
\eq
Therefore, using these new fermionic and slave-bosonic operators, the QD and tunneling Hamiltonians (\ref{qd}) and 
(\ref{tunneling}) can be rewritten, respectively, as:
\bn
H_{QD}&=&  \sum_{\sigma} \epsilon_{d\sigma}(t) f_{\sigma }^{\dagger } f_{\sigma } ,\label{qd2} \cr
H_{T} &=& \sum_{\eta k \sigma} [ V_{\eta k \sigma}(t) c_{\eta k \sigma }^{\dagger } b^{\dagger} f_{\sigma}^{\pdag} + {\rm 
{H.c.}}]. \label{tunneling2}
\en

Furthermore, as far as the three possible single electric states are considered as the basis, the expectation values of the 
diagonal elements of the density matrix, $\rho_{\alpha\alpha}$ ($\alpha=\{0, \sigma\}$), give the occupation probabilities of the 
resonant level in the QD being empty, or singly occupied by spin-$\sigma$ electron, respectively. In the slave particle notation, 
the corresponding relations between the density matrix elements and these auxiliary operators are obvious as $\hat 
\rho_{00}=|0\rangle \langle 0 |=b^{\dagger} b, \, \hat \rho_{\sigma \sigma}=|\sigma \rangle \langle \sigma |=f_{\sigma}^{\dagger} 
f_{\sigma}^{\pdag}$. According to Eq.~(\ref{comp2}), we have
\bq
\rho_{00}+\sum_{\sigma} \rho_{\sigma \sigma}=1. \label{norm}
\eq

\section{Theoretical Methods}

In this section, we derive the second-order quantum rate equations (SOQREs) to describe the higher-order tunneling processes of 
an interacting QD modeled by Eq.~(\ref{ham}) subject to arbitrary time-dependent potentials.
Since one of the most important observables in the transport problem are the electron occupation number $\rho_{\alpha\alpha}(t)$ 
in the central region, we start from the Heisenberg EOM of the density matrix element:
\bn
i\dot \rho_{\sigma\sigma}(t)&=&\langle [f_{\sigma}^\dagger f_{\sigma}(t) , H(t)] \rangle \cr
&=& \sum_{\eta k} V_{\eta k \sigma}(t) \left [ \langle f_{\sigma}^\dagger b(t) c_{\eta k \sigma}(t) - c_{\eta k 
\sigma}^\dagger(t) b^\dagger f_{\sigma}(t) \rangle \right ]. \cr
&&\label{heisenberg}
\en
Another important observable is the current $I_{\eta}(t)$ from one of the electrodes to the central region under nonequilibrium 
condition, which is defined as the time evolution of the occupation number operator of the given lead:
\bq
I_{\eta}(t)=-e\langle \dot N_{\eta}\rangle=-\frac{ie}{\hbar} \langle [H, N_{\eta}]\rangle, \nonumber
\eq
with $N_{\eta}=\sum_{k \sigma} c_{\eta k \sigma}^\dagger c_{\eta k \sigma}$. Since $H_C$ and $H_{QD}$ commute with $N_{\eta}$, 
one readily finds
\bq
I_{\eta}(t)=\frac{ie}{\hbar} \sum_{k \sigma} V_{\eta k \sigma}(t) \left [ \langle c_{\eta k \sigma}^\dagger(t) b^\dagger 
f_{\sigma}(t) - f_{\sigma}^\dagger b(t) c_{\eta k \sigma}(t) \rangle \right ]. \label{currentformula}
\eq

By defining the first-order ADMEs, the current matrices, $\hat\Pi_{\eta\sigma}(t)$, as
\bq
\hat\Pi_{\eta\sigma}(t)= i \sum_{k} V_{\eta k \sigma}(t) c_{\eta k \sigma}^\dagger(t) b^\dagger f_{\sigma}(t), \label{cm}
\eq
we can write the EOM of density matrix element Eq.~(\ref{heisenberg}) and the current Eq.~(\ref{currentformula}) in term of the 
expectation value of the current matrices, $\Pi_{\eta\sigma}(t)=\langle \hat\Pi_{\eta\sigma}(t)\rangle$, as
\bn
\dot \rho_{\sigma\sigma}(t) &=& \sum_{\eta} \left [ \Pi_{\eta\sigma}(t) + \Pi_{\eta\sigma}^\dagger(t) \right ],  \label{dotrho} 
\\
I_{\eta}(t) &=& \frac{e}{\hbar} \sum_{\sigma} \left [ \Pi_{\eta\sigma}(t) + \Pi_{\eta\sigma}^\dagger \right ]. \label{current}
\en
Therefor, one has to calculate the current matrices by solving its own EOM, which will unavoidably cause some new unknown 
observables, second-order ADMEs, due to the existence of Coulomb interaction in the central region. This procedure will generate 
an infinite hierarchy of coupled EOMs for these ADMEs. In order to form a closed set of equations, some suitable decoupling 
approximations have to be invoked to break this hierarchy at certain level. The choice of an appropriate truncation procedure is 
critical in order to treat properly the strong correlation effects both from the on-site Coulomb interaction and from the 
dot-lead tunnel couplings. Several years ago, a SOvN approach was derived for electronic coherent transport through an 
interacting QD by only considering two-electron coherent tunneling processes and taking the Markov limit for the two-electron 
transition events.\cite{Pedersen} This SOvN approach is believed to be in full agreement with the method of the diagrammatic 
real-time perturbation theory and to be able to describe correlated two-electron tunneling, cotunneling, because the current is 
calculated to higher order in the lead-dot coupling in this approach. However, the SOvN approach only works for temperatures 
above the Kondo temperature and gives incorrect time evolution behaviors of the tunneling current and electron occupancies. In 
this paper, we want to overcome the two drawbacks. First, a systematic and physically more clear decoupling approximation is 
introduced, e.g. the famous Lacroix's decoupling approximation,\cite{Lacroix} to tackle the electronic correlation. Especially 
different from the SOvN approach, the present SOQREs will contain the back-action effect of the kinetics of the QD to the 
electrodes. Second, our approach allows for easy incorporation for arbitrary time-dependent effects in higher order quantum 
transport through an interacting QD. For these purposes, we will, in the following, generalize the recently developed TDNGF 
method with auxiliary-mode expansion to the interacting case.

\subsection{Current matrices and its auxiliary-mode expansion}

The key quantity is here the current matrices defined in Eq.~(\ref{cm}). By applying the TDNGF with the Langreth's analytic 
continuation rules, its expectation value can be expressed in terms of the Keldysh GFs of the QD as
\bn
\Pi_{\eta\sigma}(t)&=&\sum_{k} V_{\eta k \sigma}(t) G_{\sigma,\eta k \sigma}^<(t,t) \cr
&=& \sum_{k} V_{\eta k \sigma}(t) \int_{-\infty}^t dt_1 V_{\eta k \sigma}^*(t_1) \left [ G_{\sigma}^r(t,t_1) \right. \cr
&& \left. \times g_{\eta k \sigma}^<(t_1,t) + G_{\sigma}^<(t,t_1) g_{\eta k \sigma}^a(t_1,t) \right ] \cr
&=& \int_{-\infty}^t dt_1 \left [ G_{\sigma}^>(t,t_1) \Sigma_{\eta \sigma}^<(t_1,t) \right. \cr
&& \left. - G_{\sigma}^<(t,t_1) \Sigma_{\eta \sigma}^>(t_1,t) \right ]. \label{cm2}
\en
The superscripts $<,>$ denote the lesser or greater indices and $r,a$ denote the retarded or advanced indices of these two-time 
functions. $G_{\sigma, \eta k \sigma}(t,t')$ and $G_{\sigma}(t,t')$ are the mixture GF and the GFs of the QD, respectively, which 
are defined using Zubarev's notation as: $G_{\sigma, \eta k \sigma}^\nu(t,t')= \langle\langle b^\dagger f_{\sigma}(t)| c_{\eta k 
\sigma}^\dagger (t')\rangle\rangle^\nu$, and $G_{\sigma}^\nu(t,t') =\langle\langle b^\dagger f_{\sigma}(t)| f_{\sigma}^\dagger b 
(t') \rangle\rangle^\nu$ ($\nu=\{<,>,r,a\}$). $g_{\eta k \sigma}(t,t')$ are the exact time-dependent GFs in the leads for the 
uncoupled system:
\bn
g_{\eta k\sigma}^{<}(t,t')&\equiv& i\langle c_{\eta k\sigma}^\dagger(t') c_{\eta k\sigma}(t) \rangle \cr
&=& i f_\eta(\epsilon_{\eta k}^0) \exp \left [ -i \int_{t'}^t d\tau \epsilon_{\eta k}(\tau) \right ] , \\
g_{\eta k\sigma}^{>}(t,t')&\equiv& -i\langle c_{\eta k\sigma}(t) c_{\eta k\sigma}^\dagger(t') \rangle \cr
&=& -i \left [1-f_\eta(\epsilon_{\eta k}^0) \right] \exp \left [ -i \int_{t'}^t d\tau \epsilon_{\eta k}(\tau) \right ] , \cr
&& \\
g_{\eta k\sigma}^{r,a}(t,t') &\equiv& \mp i \theta(\pm t \mp t') \langle \{c_{\eta k\sigma}(t), c_{\eta k\sigma}^\dagger(t') 
\}\rangle \cr
&=& \mp i \theta(\pm t \mp t') \exp \left [ -i \int_{t'}^t d\tau \epsilon_{\eta k}(\tau) \right ] ,
\en
where $f_{\eta}(\varepsilon)$ is the Fermi distribution function of the lead $\eta$ with the chemical potential $\mu_{\eta}$ and 
the temperature $T$ [$\beta=(k_B T)^{-1}$],
$$
f_{\eta}(\varepsilon)=\frac{1}{1+e^{\beta (\varepsilon-\mu_{\eta})}}.
$$
$\Sigma_{\eta\sigma}^{<(>)}(t_1,t)$ are the corresponding self-energies of the QD due to tunnel-coupling to electrode $\eta$:
\bn
\Sigma_{\eta \sigma}^<(t_1,t)&=& \sum_{k} V_{\eta k \sigma}^*(t_1) g_{\eta k\sigma}^<(t_1,t) V_{\eta k \sigma}(t) \cr
&=& i\int\frac{d\varepsilon}{2\pi} f_{\eta}(\varepsilon) e^{-i\varepsilon(t_1-t)} \Gamma_{\eta \sigma}(\varepsilon,t_1,t), 
\label{selfen1} \\
\Sigma_{\eta \sigma}^>(t_1,t)&=& \sum_{k} V_{\eta k \sigma}^*(t_1) g_{\eta k\sigma}^>(t_1,t) V_{\eta k \sigma}(t) \cr
&=& -i\int\frac{d\varepsilon}{2\pi} f_{\eta}^-(\varepsilon) e^{-i\varepsilon(t_1-t)} \Gamma_{\eta \sigma}(\varepsilon,t_1,t), 
\label{selfen2}
\en
with $f_{\eta}^-(\varepsilon)=1-f_{\eta}(\varepsilon)$. The generalized level-width function $\Gamma_{\eta}(\varepsilon,t,t')$ 
depends on the density of states $\varrho_{\eta}(\varepsilon)$ of lead $\eta$ and the coupling $V_{\eta 
\sigma}(\varepsilon,t)=V_{\eta k \sigma}(t)$
\bn
\Gamma_{\eta \sigma}(\varepsilon,t_1,t)&=& 2\pi \varrho_{\eta}(\varepsilon) V_{\eta \sigma}(\varepsilon,t) V_{\eta 
\sigma}^*(\varepsilon,t_1) \cr
&& \times \exp \left [ i\int_{t_1}^{t} d\tau \Delta_{\eta}(\tau)\right ]. \label{glf}
\en
To obtain the third equality in Eq.~(\ref{cm2}), we employ the relation for the two-time functions, the Green's functions and the 
self-energies,  defined on the time forward and backward branches:
\bq
X^{r,a}(t,t')= \pm \theta(\pm t \mp t') \left [ X^>(t,t') - X^<(t,t') \right ]. \label{ralg}
\eq

For further calculation of the self-energies, the detail information about the linewidth function containing the density of 
states $\varrho_{\eta}(\varepsilon)$ and the energy-dependent tunneling matrix element $V_{\eta \sigma}(\varepsilon)$ have to be 
given as a input from the realistic first-principles calculation or be set as a specific toy model. Actually, two theoretical 
models are usually used in the context of quantum transport problem. One is the Lorentzian-form linewidth function,
\bq
\Gamma_{\eta \sigma}(\varepsilon)=2\pi \varrho_{\eta}(\varepsilon) V_{\eta \sigma}(\varepsilon) V_{\eta \sigma}^*(\varepsilon) 
=\frac{\Gamma_{\eta \sigma}^0W^2}{\varepsilon^2+W^2},
\eq
with the linewidth amplitude $\Gamma_{\eta \sigma}^0$ andd the bandwidth $W$. Lorentzian linewidth is believed to provide a 
mathematically convenient way to introduce finite-bandwidth effects on electronic transport through nanoscale device. The other 
simpler one is the so-called wideband limit (WBL), which is valid when the density of states in the leads varies slowly with 
energy in the vicinity of the levels of the central region. In this limit, the density of states of the lead $\eta$, 
$\varrho_{\eta}$, and the coupling between the leads and the central region, $V_{\eta \sigma}^0$, are both assumed to be constant 
in the energy window relevant for transport. Obviously, this results in a tremendous simplification of mathematical calculation. 
Here, in order to concentrate our attention on the main objective of this paper, the correlation effect on higher order 
electronic tunneling through a QD, we chose the WBL in the following calculation. Therefore, the generalized linewidth function 
Eq.~(\ref{glf}) becomes energy independence, $\Gamma_{\eta \sigma}(\varepsilon,t_1,t)=\Gamma_{\eta \sigma}(t_1,t)$ and
\bq
\Gamma_{\eta \sigma}(t_1,t)=\Gamma_{\eta \sigma}^0 u_\eta(t) u_\eta^*(t_1) \exp \left [ i\int_{t_1}^{t} d\tau 
\Delta_{\eta}(\tau)\right ], \label{slf}
\eq
with a constant tunnel-coupling strength $\Gamma_{\eta \sigma}^0 = 2\pi \sum_{k} |V_{\eta k \sigma}^0|^2 
\delta(\varepsilon-\epsilon_{\eta k}^0)$. Nevertheless, it should be noted that our methodology is also applicable if the 
Lorentzian linewidth function is utilized.

Moreover, the next key skill is to derive analytical expressions for the self-energies Eqs.~(\ref{selfen1}) and (\ref{selfen2}), 
i.e., the integral kernel of the current matrices in Eq.~(\ref{cm2}), by evaluating integrals involving the Fermi distribution  
function. Such expressions will needless to say significantly cut down the heavy computational burden for the simulations of 
transient response. In literature, the Matsubara expansion of the Fermi function has usually been applied for energy integrals by 
means of contour integration using the residue theorem. However, its slow convergence makes it not suitable for numerical 
applications in fermionic many-body theory and time evolution problems in the present investigation. Recently, several highly 
accurate and much more efficient approximations have been proposed for the Fermi function, such as the continued fraction 
representation,\cite{Ozaki} the partial fractional decomposition,\cite{Zheng2,Croy3} and the Pad\'e spectral 
decomposition.\cite{Xie,Karrasch,Hu} All of these decompositions have similar mathematical structure with the usual Matsubara 
expansion, that is an infinite summation over simple poles which are all located on the imaginary axis. As a result, these 
decompositions can be easily incorporated into many applications with the same precedure as for the Matsubara summation. In 
particular, the partial fractional decomposition was recently used by Croy and Saalmann to transform the integration formula of 
the current matrices, Eq.~(\ref{cm2}), into the summation form, the auxiliary-mode expansion as is called by them, in the context 
of time dependent electron transport in a noninteracting QD. Here we will use one more efficient approximation, the Pad\'e 
approximation, to expand the Fermi function as follows:\cite{Karrasch,Hu}
\bq
f_{\eta}(\varepsilon) \simeq \frac{1}{2} - \sum_{p=1}^{N_p} \frac{R_{p}}{\beta} \left ( \frac{1}{\varepsilon -\chi_{\eta p}^+} + 
\frac{1}{\varepsilon - \chi_{\eta p}^-} \right ), \label{pade}
\eq
where $R_p$ is the $p$th residue in the Pad\'e decomposition, and $\chi_{\eta p}^\pm = \mu_{\eta} \pm i \chi_{p}/\beta$, $i 
\chi_{p}$ is the $p$th
Pad\'e pole in the upper complex plane, i.e., $\chi_p>0$ (see the Appendix for the detail of how to calculate these parameters). 
$N_p$ is the number of Pad\'e pole pairs.

Substituting the energy-independent linewidth function, Eq.~(\ref{slf}), and the Pad\'e expansion of the Fermi distribution 
function, Eq.~(\ref{pade}), into the lesser and greater self-energies, Eqs.~(\ref{selfen1}) and (\ref{selfen2}), the energy 
integration can be accomplished by applying contour integration and the Cauchy's residue theorem, and these self-energies can 
thus be transformed into the summation forms
\bq
\Sigma_{\eta \sigma}^{\lessgtr}(t_1,t)= \pm i \frac{1}{2} \Gamma_{\eta \sigma}^0 |u_{\eta}(t)|^2 \delta(t_1-t) + u_{\eta}(t) 
\sum_{p} \Sigma_{\eta \sigma p}^+ (t_1,t),
\eq
with
\bq
\Sigma_{\eta \sigma p}^+ (t_1,t) = u_{\eta}^*(t_1) \frac{R_{p}}{\beta} \Gamma_{\eta \sigma}^0 e^{i\int_{t_1}^{t}d\tau \chi_{\eta 
p}^+ (\tau)},
\eq
for $t_1<t$, and
\bq
\Sigma_{\eta \sigma}^{\lessgtr}(t_1,t)= \pm i \frac{1}{2} \Gamma_{\eta \sigma}^0 |u_{\eta}(t)|^2 \delta(t_1-t) - u_{\eta}^*(t_1) 
\sum_{p} \Sigma_{\eta \sigma p}^- (t_1,t),
\eq
with
\bq
\Sigma_{\eta \sigma p}^- (t_1,t) = u_{\eta}(t) \frac{R_{p}}{\beta} \Gamma_{\eta \sigma}^0 e^{i\int_{t_1}^{t}d\tau \chi_{\eta p}^- 
(\tau)},
\eq
for $t_1>t$, respectively. Here $\chi_{\eta p}^\pm (t)= \chi_{\eta p}^\pm + \Delta_{\eta}(t)$.
Then, we insert the expanded self-energies into Eq.~(\ref{cm2}) to get
\bq
\Pi_{\eta \sigma}(t) = \frac{1}{4} \Gamma_{\eta\sigma}^0 |u_{\eta}(t)|^2 (\rho_{00} - \rho_{\sigma\sigma}) + u_{\eta}(t) \sum_{p} 
\Pi_{\eta \sigma p}(t), \label{cm3}
\eq
and an {\em auxiliary-mode expansion} of the current matrices, $\Pi_{\eta \sigma p}(t)$,
\bn
\Pi_{\eta \sigma p}(t) &=& \int_{-\infty}^t dt_1 \Sigma_{\eta \sigma p}^+(t_1,t) \left [ G_{\sigma}^>(t,t_1) - 
G_{\sigma}^<(t,t_1) \right ] \cr
&=&  \int_{-\infty}^t dt_1 \Sigma_{\eta \sigma p}^+(t_1,t) \left [ G_{\sigma}^r(t,t_1) - G_{\sigma}^a(t,t_1) \right]. \cr
&& \label{cmame}
\en
To obtain Eq.~(\ref{cm3}), we use $G_{\sigma}^<(t,t)=i\rho_{\sigma\sigma}$ and $G_{\sigma}^>(t,t)=-i\rho_{00}$ according to the 
definition of the contour-ordered GFs. Likewise, it is easy to find
\bq
\Pi_{\eta \sigma}^\dagger (t) = \frac{1}{4} \Gamma_{\eta\sigma}^0 |u_{\eta}(t)|^2 (\rho_{00} - \rho_{\sigma\sigma}) + 
u_{\eta}^*(t) \sum_{p} \Pi_{\eta \sigma p}^*(t), \label{cm4}
\eq
and
\bq
\Pi_{\eta \sigma p}^*(t) = \int_{-\infty}^t dt_1 \Sigma_{\eta \sigma p}^-(t,t_1) \left [ G_{\sigma}^<(t_1,t) - 
G_{\sigma}^>(t_1,t) \right ],
\eq
because of $[G_{\sigma}^{\pm}(t,t')]^\dagger=-G_{\sigma}^{\pm}(t',t)$ and $(\chi_{\eta p}^-)^*=\chi_{\eta p}^+$.

\subsection{Equations of motion for $\Pi_{\eta\sigma p}$ and the second-order ADME}

Performing time derivative of Eq.~(\ref{cmame}) yields
\bn
i\frac{\partial}{\partial t} \Pi_{\eta\sigma p}(t) &=& u_{\eta}^*(t) \frac{R_p}{\beta} \Gamma_{\eta \sigma}^0 (\rho_{00} + 
\rho_{\sigma\sigma}) - \chi_{\eta p}^+(t) \Pi_{\eta\sigma p}(t) \cr
&& \hspace{-1.5cm} + \int_{-\infty}^t dt_1 \Sigma_{\eta \sigma p}^+(t_1,t) i\frac{\partial}{\partial t} \left [ 
G_{\sigma}^r(t,t_1) - G_{\sigma}^a(t,t_1) \right ]. \label{eomcmame}
\en
This equation shows that the EOM for the auxiliary-mode expansion of the current matrices relies on the EOMs for the retarded and 
advance GFs of the QD,
\bn
i\frac{\partial}{\partial t}G_{\sigma}^{r(a)}(t,t_1) &=& \delta(t-t_1) \langle \{ b^\dagger f_{\sigma}(t), f_{\sigma}^\dagger 
b(t) \}\rangle \cr
&& + \langle\langle \left [ b^\dagger f_{\sigma}(t), H(t) \right ] | f_{\sigma}^\dagger b(t_1) \rangle\rangle^{r(a)} \cr
&& \hspace{-2cm} = [\rho_{00}(t)+\rho_{\sigma\sigma}(t)] \delta(t-t_1) + \epsilon_{d\sigma}(t) G_{\sigma}^{r(a)}(t,t_1) \cr
&& \hspace{-2cm}+ \sum_{\eta' k'} \left \{ V_{\eta' k' \sigma}^*(t) \left [ G_{\eta' k'\sigma,\sigma}^{r(a)}(t,t_1) - G_{\eta' 
k'\sigma,\sigma}^{\bar\sigma\bar\sigma, r(a)}(t,t_1) \right] \right. \cr
&& \left. + V_{\eta' k' \bar\sigma}^*(t) G_{\eta' k'\bar\sigma,\sigma}^{\bar\sigma\sigma, r(a)}(t,t_1) \right \},  \label{eomgfr}
\en
where three new GFs appear in the last equality: the mixture GF $G_{\eta' k'\sigma,\sigma}^\nu(t,t')=\langle\langle c_{\eta' 
k'\sigma}(t)| f_{\sigma}^\dagger b(t')\rangle\rangle^\nu$ due to coupling to electrodes, and the two two-particle mixture GF 
$G_{\eta' k'\sigma_1,\sigma}^{\sigma_2\sigma_3, \nu}(t,t')=\langle\langle f_{\sigma_2}^\dagger(t) f_{\sigma_3}(t) c_{\eta' 
k'\sigma_1}(t)|f_{\sigma}^\dagger b(t')\rangle\rangle^\nu$ generated by the on-site Coulomb interaction inside the QD.
Following standard EOM analysis, one can easily derive the Dyson equation for the mixture GF,
\bq
G_{\eta' k'\sigma,\sigma}^{r(a)}(t,t_1)= \int dt_2 g_{\eta' k'\sigma}^{r(a)}(t,t_2) V_{\eta' k' \sigma}(t_2) 
G_{\sigma}^{r(a)}(t_2,t_1). \label{gfmix}
\eq
Substituting Eq.~(\ref{eomgfr}) into the equation Eq.~(\ref{eomcmame}) and applying Eq.~(\ref{gfmix}), we get
\bn
i\frac{\partial}{\partial t} \Pi_{\eta\sigma p}(t) &=& u_{\eta}^*(t) \frac{R_p}{\beta} \Gamma_{\eta \sigma}^0 (\rho_{00} + 
\rho_{\sigma\sigma}) \cr
&& \hspace{-1cm} + \left [ \epsilon_{d\sigma}(t) - \frac{i}{2} \Gamma_{\sigma}(t) - \chi_{\eta p}^+(t) \right ] \Pi_{\eta\sigma 
p}(t) \cr
&& \hspace{-1cm} + \sum_{\eta'} \left [ \Omega_{\eta\sigma p, \eta'1}(t)- \Omega_{\eta\sigma p, \eta'2}(t) \right ], 
\label{eomcmame2}
\en
where $\Gamma_{\sigma}(t)=\sum_{\eta'} |u_{\eta'}(t)|^2 \Gamma_{\eta'\sigma}^0$, and two new second-order ADMEs 
$\Omega_{\eta\sigma p, \eta'\alpha}(t)$ ($\alpha=1,2$) are introduced as
\bn
\Omega_{\eta\sigma p, \eta'1}(t) &=& \int_{-\infty}^t dt_1 \Sigma_{\eta \sigma p}^+(t_1,t) \sum_{k'} V_{\eta' k'\bar\sigma}^*(t) 
\cr
&&\hspace{-1cm}\times \left [ G_{\eta' k' \bar\sigma,\sigma}^{\bar\sigma\sigma, r}(t,t_1) -G_{\eta' k' \bar\sigma, 
\sigma}^{\bar\sigma\sigma, a}(t,t_1) \right], \label{omega1} \\
\Omega_{\eta\sigma p, \eta'2}(t) &=& \int_{-\infty}^t dt_1 \Sigma_{\eta \sigma p}^+(t_1,t) \sum_{k'} V_{\eta' k'\sigma}^*(t) \cr
&&\hspace{-1cm}\times \left [ G_{\eta' k' \sigma,\sigma}^{\bar\sigma\bar\sigma, r}(t,t_1) -G_{\eta' k' 
\sigma,\sigma}^{\bar\sigma\bar\sigma, a}(t,t_1) \right]. \label{omega2}
\en

For the case of noninteracting system, the second-order ADMEs are vanishing in the WBL, and thus the equations (\ref{dotrho}), 
(\ref{cm3}), (\ref{cm4}), and (\ref{eomcmame2}) together with the normalization relation Eq.~(\ref{norm}) already constitute the 
closed set of coupled EOMs, which are exactly the same as those derived by Croy and Saalmann.\cite{Croy1} While for an 
interacting QD under present study, one needs to make some approximations to calculate the second order GFs.

\subsubsection{Hubbard-I approximation}

First, we employ the Hartree-Fock approximation for the second order GF:
\bn
G_{\eta' k'\sigma_1,\sigma}^{\sigma_2\sigma_3, \nu}(t,t_1) &=& \langle\langle f_{\sigma_2}^\dagger(t) f_{\sigma_3}(t) c_{\eta' 
k'\sigma_1}(t)|f_{\sigma}^\dagger b(t')\rangle\rangle^\nu \cr
&\approx& \langle f_{\sigma_2}^\dagger(t) f_{\sigma_3}(t)\rangle \langle\langle c_{\eta' k'\sigma_1}(t)|f_{\sigma}^\dagger 
b(t')\rangle\rangle^\nu \cr
&=& \delta_{\sigma_2\sigma_3} \rho_{\sigma_2\sigma_2}(t) \delta_{\sigma_1\sigma}G_{\eta' k' \sigma,\sigma}^\nu(t,t_1). \nonumber
\en
In this case, we can readily obtain the retarded GF of the QD, from Eq.~(\ref{eomgfr}), in the frequency domain,
\bq
G_{\sigma}^{r}(\varepsilon)=\frac{1-\rho_{\bar \sigma \bar \sigma}}{\varepsilon-\epsilon_{0\sigma} + \frac{i}{2}\Gamma_{\sigma} 
(1-\rho_{\bar \sigma \bar \sigma})}, \label{gfrhia}
\eq
for the time-independent situation. This is usually referred to as the Hubbard-I approximation (HIA) in the context of the GF 
theory for strongly correlated electron systems. Thus, the EOM for the current matrices is also closed and takes the form
\bn
i\frac{\partial}{\partial t} \Pi_{\eta\sigma p}(t) &=& u_{\eta}^*(t) \frac{R_p}{\beta} \Gamma_{\eta \sigma}^0 (1- 
\rho_{\bar\sigma\bar\sigma}) \cr
&& \hspace{-1cm} + \left [ \epsilon_{d\sigma}(t) - \frac{i}{2} \Gamma_{\sigma}'(t) - \chi_{\eta p}^+(t) \right ] \Pi_{\eta\sigma 
p}(t) , \label{eomcmame3}
\en
with $\Gamma_{\sigma}'(t)=\sum_{\eta'} |u_{\eta'}(t)|^2 (1-\rho_{\bar\sigma \bar\sigma}(t))\Gamma_{\eta'\sigma}^0$. Worth 
noticing that the resulting EOM Eq.~(\ref{eomcmame3}) is similar with that derived by using EOM for operators, except for the 
extra factor $1-\rho_{\bar\sigma\bar\sigma}(t)$ in $\Gamma_{\sigma}'(t)$ in our result. Removing the factor 
$1-\rho_{\bar\sigma\bar\sigma}$ and spin indices in Eq.~(\ref{eomcmame3}), this equation reduces exactly to the previously 
developed EOM for the current matrices of the noninteracting single-level QD system in the WBL.\cite{Croy1}

It is known that the HIA is reasonable at relatively high temperatures, where the Kondo resonance does not appear during electron 
tunneling processes, at $T\gg T_{K}$ ($T_{K}$ is the Kondo temperature). However, to investigate the system at low temperatures, 
where the higher order coherent tunneling (cotunneling) is the dominating transport mechanism involving two-electron correlated 
scattering processes, and even the Kondo effect plays a certain role, it is necessary to calculate the second order GF beyond 
this level of  approximation.

\subsubsection{Lacroix's decoupling approximation}

At first, we derive the EOMs of the second order GFs, which will generate three third order GFs,
\bn
i\frac{\partial}{\partial t}G_{\eta' k'\bar\sigma,\sigma}^{\bar\sigma\sigma, r}(t,t_1) &=& -\langle f_{\bar\sigma}^\dagger b 
c_{\eta'k'\bar\sigma}(t) \rangle \delta(t-t_1) \cr
&& \hspace{-3cm} + [\epsilon_{\eta'k'}(t) + \epsilon_{d\sigma}(t) - \epsilon_{d\bar\sigma}(t) ] G_{\eta'k'\bar\sigma, 
\sigma}^{\bar\sigma\sigma,r}(t,t_1) \cr
&& \hspace{-3cm} + \sum_{\eta'' k''} V_{\eta'' k''\bar\sigma}(t) \langle\langle c_{\eta''k'' \bar\sigma}^\dagger (t) c_{\eta' k' 
\bar\sigma}(t) b^\dagger f_{\sigma}(t) | f_{\sigma}^\dagger b(t_1) \rangle\rangle^r \cr
&& \hspace{-3cm} - \sum_{\eta'' k''} V_{\eta'' k''\sigma}^*(t) \langle\langle f_{\bar\sigma}^\dagger b(t) c_{\eta' k' 
\bar\sigma}(t) c_{\eta''k'' \sigma}(t) | f_{\sigma}^\dagger b(t_1)\rangle\rangle^r, \nonumber
\en
\bn
i\frac{\partial}{\partial t}G_{\eta' k'\sigma,\sigma}^{\bar\sigma\bar\sigma, r}(t,t_1) &=& \epsilon_{\eta' k'}(t) G_{\eta' k' 
\sigma}^{\bar\sigma\bar\sigma, r}(t,t_1) \cr
&& \hspace{-3cm} + \sum_{\eta'' k''} V_{\eta'' k''\bar\sigma}(t) \langle\langle c_{\eta''k'' \bar\sigma}^\dagger (t) c_{\eta' k' 
\sigma}(t) b^\dagger f_{\bar\sigma}(t) | f_{\sigma}^\dagger b(t_1)\rangle\rangle^r \cr
&& \hspace{-3cm} + \sum_{\eta'' k''} V_{\eta'' k''\bar\sigma}^*(t) \langle\langle f_{\bar\sigma}^\dagger b(t) c_{\eta''k'' 
\bar\sigma}(t) c_{\eta' k' \sigma}(t) | f_{\sigma}^\dagger b(t_1)\rangle\rangle^r. \nonumber
\en
Notice that all these next generation of GFs involve the excitation of two electrons in the electrodes. In what follow, we employ 
the well-known truncated scheme, the Lacroix's decoupling approximation (LDA) to contract pairs of same-spin lead-lead electron 
operators and pairs of same-spin dot-lead electron operators:\cite{Lacroix}
\bn
\langle\langle c_{\eta''k'' \bar\sigma}^\dagger (t) c_{\eta' k' \bar\sigma}(t) b^\dagger f_{\sigma}(t) | f_{\sigma}^\dagger 
b(t_1) \rangle\rangle^\nu &\approx& \cr
&& \hspace{-3cm} \langle c_{\eta''k'' \bar\sigma}^\dagger c_{\eta' k' \bar\sigma}(t) \rangle G_{\sigma}^\nu(t,t_1), \cr
\langle\langle f_{\bar\sigma}^\dagger b(t) c_{\eta' k' \bar\sigma}(t) c_{\eta''k'' \sigma}(t) | f_{\sigma}^\dagger 
b(t_1)\rangle\rangle^\nu &\approx& \cr
&& \hspace{-3cm} \langle f_{\bar\sigma}^\dagger b c_{\eta' k' \bar\sigma}(t) \rangle G_{\eta''k''\sigma,\sigma}^\nu(t,t_1), \cr 
\langle\langle c_{\eta''k'' \bar\sigma}^\dagger (t) c_{\eta' k' \sigma}(t) b^\dagger f_{\bar\sigma}(t) | f_{\sigma}^\dagger 
b(t_1)\rangle\rangle^\nu &\approx& \cr
&& \hspace{-3cm} -\langle c_{\eta''k'' \bar\sigma}^\dagger b^\dagger f_{\bar\sigma}(t) \rangle 
G_{\eta'k'\sigma,\sigma}^\nu(t,t_1). \nonumber
\en
Then the formal solutions of the second generation of GFs are
\begin{widetext}
\bn
G_{\eta' k'\bar\sigma,\sigma}^{\bar\sigma\sigma, r}(t,t_1) &=& -\langle f_{\bar\sigma}^\dagger b c_{\eta'k'\bar\sigma}(t_1) 
\rangle \widetilde{g}_{\eta'k'\bar\sigma}^{r}(t,t_1) + \sum_{\eta''k''}\int dt_2  \widetilde{g}_{\eta'k'\bar\sigma}^{r}(t,t_2) 
V_{\eta''k''\bar\sigma}(t_2) \langle c_{\eta''k'' \bar\sigma}^\dagger  c_{\eta' k' \bar\sigma}(t_2) \rangle G_{\sigma}^r(t_2,t_1) 
\cr
&&  - \sum_{\eta''k''} \int dt_2 \widetilde{g}_{\eta'k'\bar\sigma}^{r}(t,t_2) V_{\eta''k''\sigma}^*(t_2) \langle 
f_{\bar\sigma}^\dagger b c_{\eta' k' \bar\sigma}(t_2) \rangle G_{\eta''k''\sigma,\sigma}^r(t_2,t_1), \label{LA} \\
G_{\eta' k'\sigma,\sigma}^{\bar\sigma\bar\sigma, r}(t,t_1) &=& -i \int dt_2 g_{\eta'k'\sigma}^r(t,t_2) I_{\bar\sigma}(t_2) 
G_{\eta'k'\sigma,\sigma}^{r}(t_2,t_1) ,
\en
where $I_{\bar\sigma}(t)= \sum_{\eta}[\Pi_{\eta\bar\sigma}(t)+\Pi_{\eta\bar\sigma}^\dagger(t)]$ is the spin-resolved current 
difference between the left and right leads, and the auxiliary two-time retarded function $\widetilde{g}_{\eta k 
\bar\sigma}^r(t,t')$ is defined as
\bq
\left (i\frac{\partial}{\partial t} - \epsilon_{\eta k}(t) - \delta_{d\sigma}(t) \right) \widetilde{g}_{\eta k 
\bar\sigma}^r(t,t') = \delta(t-t'),
\eq
with $\delta_{d\sigma}(t)=\epsilon_{d\sigma}(t)-\epsilon_{d\bar\sigma}(t)$.
Inserting the two formal solutions into Eqs.~(\ref{omega1}) and (\ref{omega2}) and after lengthy and cumbersome calculations, we 
can derive closed explicit expressions for the second-order ADMEs, under this decoupling approximation,
\bn
\Omega_{\eta\sigma p, \eta'1}(t) &=& - \frac{i}{4} \Gamma_{\eta'\bar\sigma}^0 |u_{\eta'}(t)|^2 \Pi_{\eta \sigma p}(t) + \sum_{p'} 
u_{\eta'}^*(t) \int_{-\infty}^t dt_2 \left ( \left \{ \Gamma_{\eta'\bar\sigma}^0  \frac{R_{p'}}{\beta} u_{\eta'}(t_2) + 
\frac{i}{2} \left [ \Gamma_{\sigma}(t_2) + \Gamma_{\bar\sigma}(t_2) \right] \Pi_{\eta'\bar\sigma p'}^*(t_2) \right \} \right. \cr
&& \left. \times \Pi_{\eta \sigma p}(t_2) - \Gamma_{\eta \sigma}^0 \frac{R_{p}}{\beta} u_{\eta}^*(t_2) \Pi_{\eta'\bar\sigma 
p'}^*(t_2) \right) \exp \left \{ i\int_{t_2}^t d\tau \left [ \chi_{\eta p}^+(\tau) - \chi_{\eta' p'}^-(\tau) - 
\delta_{d\sigma}(\tau) \right ] \right \},  \label{omega1a} \\
\Omega_{\eta\sigma p, \eta'2}(t) &=& 0. \label{omega2a}
\en
\end{widetext}
Employing the two closed second-order ADMEs, we rewrite Eq.~(\ref{eomcmame2}) as
\bn
i\frac{\partial}{\partial t} \Pi_{\eta\sigma p}(t) &=& u_{\eta}^*(t) \frac{R_p}{\beta} \Gamma_{\eta \sigma}^0 (\rho_{00} + 
\rho_{\sigma\sigma}) \cr
&& \hspace{-2cm} + \left [ \epsilon_{d\sigma}(t) - \frac{i}{4} (2\Gamma_{\sigma}(t)+ \Gamma_{\bar\sigma}(t)) - \chi_{\eta p}^+(t) 
\right ] \Pi_{\eta\sigma p}(t) \cr
&& \hspace{-2cm} + \sum_{\eta'p'} u_{\eta'}^*(t) \Omega_{\eta\sigma p, \eta'\bar\sigma p'}(t), \label{eomcmame4}
\en
and the EOM for the auxiliary mode expansion of the second-order ADME, $\Omega_{\eta\sigma p, \eta'\bar\sigma p'}(t)$, reads
\bn
i\frac{\partial}{\partial t} \Omega_{\eta\sigma p, \eta'\bar\sigma p'}(t) &=& i\left \{ \Gamma_{\eta'\bar\sigma}^0  
\frac{R_{p'}}{\beta} u_{\eta'}(t) \right . \cr
&& \hspace{-2.5cm} \left. + \frac{i}{2} \left [ \Gamma_{\sigma}(t) + \Gamma_{\bar\sigma}(t) \right] \Pi_{\eta'\bar\sigma p'}^*(t) 
\right \}\Pi_{\eta \sigma p}(t) \cr
&& \hspace{-2.5cm} - i \Gamma_{\eta \sigma}^0 \frac{R_{p}}{\beta} u_{\eta}^*(t) \Pi_{\eta'\bar\sigma p'}^*(t) \cr
&& \hspace{-2.5cm} -\left [ \chi_{\eta p}^+(t) - \chi_{\eta' p'}^-(t) - \delta_{d\sigma}(t) \right ] \Omega_{\eta\sigma p, 
\eta'\bar\sigma p'}(t). \label{eomsoadme}
\en
The two explicit expressions, Eqs.~(\ref{eomcmame4}) and (\ref{eomsoadme}), for the EOMs for the first- and second-order ADMEs, 
are the main results of this paper. The two equations adding Eqs.~(\ref{dotrho}), (\ref{cm3}), (\ref{cm4}), together with the 
normalization relation Eq.~(\ref{norm}), thereby constitute the closed set of differential equations. We call them the SOQREs. 
Solving these differential equations for a given initial condition, will provide a suitable description of transient dynamics of 
higher order correlated coherent tunneling through an interacting QD under an influence of arbitrary time varying potentials.
It is interesting to notice that the dynamical equation (\ref{eomsoadme}) is nonlinear which is stemming from the on-site Coulomb 
interaction. While for the noninteracting QD, the corresponding equation is vanishing at the WBL as abovementioned or is linear 
provided that Lorentzian band width function is employed.

\section{Steady-state solutions}

In this section, we analyze the time-independent transport using the SOQREs and compare with other methods.

\subsection{Hubbard-I approximation}

First we discuss the HIA. The steady-state solution for the current matrices can be easily obtained from Eq.~(\ref{eomcmame3})
\bq
\Pi_{\eta \sigma p}= -\frac{\frac{R_p}{\beta} \Gamma_{\eta \sigma}^0 (1-\rho_{\bar\sigma\bar\sigma})}{\epsilon_{0\sigma}-  
\frac{i}{2} \Gamma_{\sigma} (1-\rho_{\bar\sigma\bar\sigma}) - \chi_{\eta p}^+}.
\eq
with $\Gamma_{\sigma}=\Gamma_{L\sigma}^0+\Gamma_{R\sigma}^0$. For spin symmetric case, we have,
\bq
\rho_{\uparrow\uparrow} = \rho_{\downarrow\downarrow} = \frac{1}{3} + \frac{4}{3\Gamma_{\sigma}} \sum_{\eta p} \Re \Pi_{\eta 
\sigma p}. \label{onhia}
\eq
This is a self-consistent equation, giving occupation number $\rho_{\sigma\sigma}$ as a function of applied bias voltage. Then 
the current of, for example, the left lead can be calculated from Eq.~(\ref{current}) as
\bn
I_{L} &=& \frac{4\Gamma_{L\sigma}^0 \Gamma_{R\sigma}^0}{\beta \Gamma_{\sigma}} \sum_{p} R_{p} \Re \left [ 
\frac{(1-\rho_{\bar\sigma\bar\sigma})}{\epsilon_{0\sigma}- \frac{i}{2} \Gamma_{\sigma} (1-\rho_{\bar\sigma\bar\sigma})-  
\chi_{Lp}^+} \right. \cr
&& \left. - \frac{(1-\rho_{\bar\sigma\bar\sigma})}{\epsilon_{0\sigma}- \frac{i}{2} \Gamma_{\sigma} 
(1-\rho_{\bar\sigma\bar\sigma})-  \chi_{Rp}^+} \right ]. \label{currenthia}
\en

On the other hand, according to the NGF approach developed by Meir and Wingreen, the electron occupation number is determined by 
solving the integral equation of the form\cite{Meir}
\bq
\rho_{\sigma\sigma}= \int \frac{d\varepsilon}{\pi} \frac{\Gamma_{L\sigma}^0 f_{L}(\varepsilon) + \Gamma_{R\sigma}^0 
f_{R}(\varepsilon)} {\Gamma_{\sigma}} \Im G_{\sigma}^r(\varepsilon),
\eq
and the current for an interacting QD is expressed by the generalized Landauer formula
\bq
I_{L} = -\int d\varepsilon \frac{4\Gamma_{L\sigma}^0 \Gamma_{R\sigma}^0}{\Gamma_{\sigma}} [f_{L}(\varepsilon) - 
f_{R}(\varepsilon)] \Im G_{\sigma}^r(\varepsilon). \label{Landauer}
\eq
Employing the analytic expression for the retarded GF under HIA of the central region, Eq.~(\ref{gfrhia}), and the P\'ade 
decomposition of the Fermi function, Eq.~(\ref{pade}), one can easily evaluate the occupation number $\rho_{\sigma\sigma}$ and 
the current $I_{L}$, which is found to be in exact agreement with the SOQRE results, Eqs.~(\ref{onhia}) and (\ref{currenthia}).

\subsection{Simplified Lacroix's decoupling approximation}

Since the SOQRE Eq.~(\ref{eomsoadme}) is a nonlinear equation, we can not obtain a simple analytical steady-state solution. About 
twenty years ago, Meir {\it et al.} developed a simplified version of the LDA (SLDA) in the EOM method of the retarded GF to 
investigate electron tunneling through a QD out of equilibrium and/or in the presence of a magnetic field.\cite{Meir} This SLDA 
produces interesting results for the bias voltage dependence of the differential conductance, showing a zero-bias anomaly and 
peak splitting due to Zeeman effect in the presence of a magnetic field. This approximation has been believed to provide 
qualitatively correct description for the Kondo-type electronic transport through a QD when the temperature is higher than the 
Kondo temperature, $T\geq T_{K}$. In their approximation, Meir {\it et al.} derived the analytical solution of the second 
retarded GF $G_{\eta'k'\sigma}^{\bar\sigma \sigma, r}(t,t_1)$ to the order of $V_{\eta k \sigma}^2$, by simplifying 
Eq.~(\ref{LA}): neglecting the inhomogeneous term involving tunneling and ignoring the back action of the electron tunneling on 
the electron distributions of electrodes, i.e., $\langle f_{\bar\sigma}^\dagger b c_{\eta'k'\bar\sigma}\rangle \approx 0$, 
$\langle c_{\eta''k'' \bar\sigma}^\dagger  c_{\eta' k' \bar\sigma}\rangle \approx \delta_{\eta''\eta'}\delta_{k''k'} 
f_{\eta'}(\epsilon_{\eta'k'}^0)$.\cite{Meir} At the end, we find that the EOM for the second-order ADME, $\Omega_{\eta\sigma p, 
\eta'\bar\sigma p'}(t)$, Eq.~(\ref{eomsoadme}) becomes a linear differential equation,
\bn
i\frac{\partial}{\partial t} \Omega_{\eta\sigma p, \eta'\bar\sigma p'}(t) &=& i \Gamma_{\eta'\bar\sigma}^0  \frac{R_{p'}}{\beta} 
u_{\eta'}(t) \Pi_{\eta \sigma p}(t) \cr
&& \hspace{-2.5cm} -\left [ \chi_{\eta p}^+(t) - \chi_{\eta' p'}^-(t) - \delta_{d\sigma}(t) \right ] \Omega_{\eta\sigma p, 
\eta'\bar\sigma p'}(t). \label{eomsoadmesla}
\en

Besides, under this approximation, an analytical expression for the retarded GF can be derived in the frequency domain as
\bq
G_{\sigma}^{r}(\varepsilon)=\frac{1-\rho_{\bar \sigma \bar \sigma}}{\varepsilon-\epsilon_{0\sigma} - \Sigma_{0\sigma}^r - 
\Sigma_{1\sigma}^r(\varepsilon)}, \label{gfrsla}
\eq
with $\Sigma_{0\sigma}^r=-i(\Gamma_{L \sigma}^0+ \Gamma_{R\sigma}^0)/2$, and
\bq
\Sigma_{1\sigma}^r(\varepsilon) = \sum_{\eta k} |V_{\eta k}^0|^2  \frac{f_{\eta}(\epsilon_{\eta k}^0)}{\varepsilon- 
\epsilon_{\eta k}^0 + i0^+}.
\eq
Since the current formula Eq.~(\ref{Landauer}) is general, we can still use it to evaluate the current through an interacting QD 
in the SLDA provided that the retarded GF is replaced by Eq.~(\ref{gfrsla}). While the SOvN approach gives a quite different 
current\cite{Pedersen}
\bq
I_{L} = -\int d\varepsilon \frac{2\Gamma_{L\sigma}^0 \Gamma_{R\sigma}^0}{|\varepsilon - \epsilon_{0\sigma} - \Sigma_{0\sigma}^r - 
\Sigma_{1\sigma}^r(\varepsilon)|^2} [f_{L}(\varepsilon) - f_{R}(\varepsilon)] .
\eq
The difference, we believe, is coming from the fact that the GF method is a double-time dynamical theory, while the SOvN is a 
single-time theory. We emphasize that our SOQRE is based on the GF method.

\subsection{Numerical discussions}

In what follows we perform numerical calculations based on the SOQREs with both the LDA and SLDA for the steady-state transport 
in absence of magnetic field. In the calculations, we only consider the symmetric system, 
$\Gamma_{L\sigma}^0=\Gamma_{R\sigma}^0=\frac{1}{2}\Gamma$, and use $\Gamma$ as energy unit throughout this paper. Moreover, we 
assume that the external bias voltage is applied only to the left lead, $\mu_L=eV$, while the chemical potential of the right 
lead remains unchanged, $\mu_R=0$.

\begin{figure}[t]
\includegraphics[height=5.5cm,width=8.5cm]{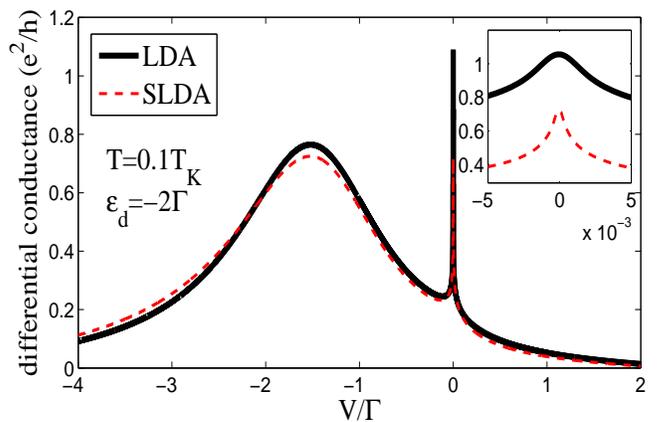}
\caption{(Colour online) The calculated steady-state differential conductance vs the applied bias voltage under $\mu_L=V$, 
$\mu_R=0$ in the SOQRE theory with LDA (solid line) and SLDA (dashed line), respectively, for the system with 
$\epsilon_d=-2\Gamma$ at the temperature $T=0.1T_K$. Inset: the detailed behavior of the Kondo peak around $V=0$.}
\label{fig1}
\end{figure}

To demonstrate the strong electron correlation effect on tunneling, we set that the energy level of the QD lies at least a 
resonance width $\Gamma$ below the chemical potentials of the leads at equilibrium. To be specific, we choose, in the following 
calculations, $\epsilon_{d}=-2\Gamma$. Therefore, this QD will show the Kondo effect, i.e., a peak at the Fermi energy in the 
equilibrium density of states (DOS), $\varrho_d(\omega=0)$, of the dot electrons, when the temperature becomes lower than a 
typical energy scale, the Kondo temperature defined as
$T_{K}=\sqrt{W\Gamma/4}\,\exp(-\pi \mid\epsilon_d\mid/\Gamma)$,
where $W$ is a high-energy cutoff equal to the half bandwidth. Here we set the half bandwidth $W=20\Gamma$, leading to the Kondo 
temperature $T_K=0.0042\Gamma$ for this QD. It is well-known that for the lead-QD-lead system, the bias-voltage-dependent 
differential conductance is proportional to the equilibrium spectral density of the dot electrons at nearly zero temperature. 
Then we first analyze the current-voltage characteristic of the system at the temperature $T=0.1T_K$ and plot the calculated 
differential conductance in Fig.~\ref{fig1}. The numerical results present two clear peaks in the DOS of the QD: a wide peak 
located nearly at $V=-2\Gamma$ corresponds to the single-particle state of the QD, and an extremely sharp peak located exactly at 
$V=0$ is stemming from the strongly correlated state at the Fermi energy, i.e., the Kondo peak. Furthermore, we find that the LDA 
gives the $\varrho_d(\omega=0)\simeq 1.05$ at $T=0.1T_K$, which is in perfect agreement with the recent analytical calculations 
based on the exact EOM solutions for the impurity's retarded GF under the same approximation.\cite{Kashcheyevs} This value is 
fifty percent larger than the SLDA's, as shown in the inset of Fig.~\ref{fig1}.

\begin{figure}[t]
\includegraphics[height=5.5cm,width=8.5cm]{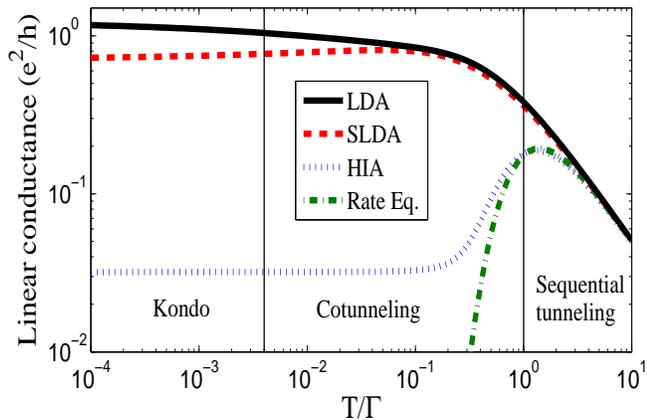}
\caption{(Colour online) The linear conductance as a function of the temperature $T/\Gamma$ varying from $10^{-4}$ to $10$ for 
the system with $\epsilon_d=-2\Gamma$. We show the calculated results by the SOQREs with three different approximations, LDA 
(solid line), SLDA (dashed line), and HIA (dashed-dotted line). For comparison, we also plot the result of the first-order rate 
equation (dotted line).}
\label{fig2}
\end{figure}

To get further comparison between the two approximations, we investigate the temperature dependences of the two-terminal linear 
conductances in a wide temperature range, from $T=10^{-4}\Gamma$ to $10\Gamma$, as shown in Fig.~\ref{fig2}. The linear 
conductance can be calculated by expanding the SOQREs to the first order with respect to the bias voltage $V$ in the limit of 
$V\rightarrow 0$. It is clear to observe that the conductance of the LDA increases gradually with decreasing of temperature up to 
nearly $1.2e^2/h$ at $T=10^{-4}\Gamma=0.025T_K$, which is also consistent with the exact EOM solutions.\cite{Kashcheyevs} 
Instead, the conductance of the SLDA exhibits an obvious decrease when the temperature becomes lower than the Kondo temperature.
These results indicate that taking full account of the inhomogeneous tunneling term and the back action effect on the electrodes 
is necessary in order to obtain qualitatively correct description of the Kondo-type transport through a QD at $T<T_K$. Therefore, 
we can conclude that the present SOQRE approach is an appropriate theoretical formulation to investigate the electronic tunneling 
in the Kondo regime and constitutes a very practical starting point for the investigation of transient Kondo dynamics.

On the contrary, the conductances of the two approximations are getting closer when the temperature is higher than the Kondo 
temperature, $T>T_K$, and becomes nearly indistinguishable at $T>0.1\Gamma$. For comparison, we also plot the linear conductances 
calculated by the HIA and the first-order rate equation\cite{Dong1,Dong2} (FORE, i.e., the lowest non-vanishing order in the 
tunneling-coupling strength $\Gamma$) in Fig.~\ref{fig2}. It is found that the HIA and the FORE are completely failure in 
describing electronic correlated tunneling through an interacting QD when the temperature is decreased to values of the order of 
the tunneling-coupling strength or lower, $T<\Gamma$. In the context of quantum transport in nanoscale system, it is well-known 
that the FORE is only valid when the temperature is higher than the tunneling strength, i.e., $T>\Gamma$, at which the sequential 
tunneling is the dominated tunneling mechanism.\cite{Schoeller,Gurvitz,Dong1} We observe actually
here that the results of the HIA and the FORE are accurately equal to those of the SOQREs at $T\geq 2\Gamma$, since the 
adequately strong thermal-broadening induces the occurrence of resonant-tunneling. From this figure, we can therefore classify 
the electronic tunneling through a single QD system into three different physical scenarios depending on the temperature: (1) At 
low temperatures, $T\leq T_K$, the strongly correlated Kondo singlet state is formed and plays a critical role in tunneling, 
leading to a great enhancement of the linear conductance. This regime is the well-known Kondo 
regime;\cite{Meir,Entin,Kashcheyevs,Roermund} (2) At the intermediate region, $T_K < T< \Gamma$, an obvious enhancement of the 
linear conductance is still observed even though the Kondo resonance is not formed. As discussed above, at this region of 
temperatures, the tunneling processes of higher order in $\Gamma$ involving two or more electrons play crucial role, which is 
called electron correlated tunneling (i.e., cotunneling) in literature.\cite{Schoeller,Dong3,Pedersen} The enhancement of 
conductance is ascribed to the renormalization effects of dot level due to coupling to the electrodes. It is reminiscence of the 
strong correlation effect; (3) At high temperatures, $T\geq \Gamma$, the QD system is in the sequential tunneling regime.

\begin{figure}[htb]
\includegraphics[height=5.5cm,width=8.5cm]{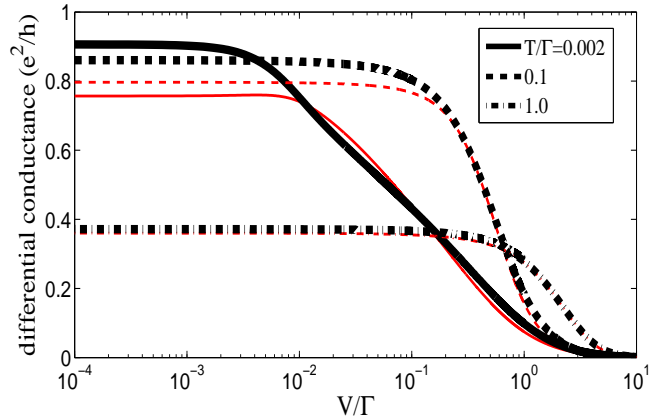}
\caption{(Colour online) The differential conductance as a function of the bias voltage for the system with $\epsilon_d=-2\Gamma$ 
at three different temperatures, $T/\Gamma=0.002$ (solid lines), $0.1$ (dashed lines), and $1.0$ (dashed-dotted lines), 
calculated by the SOQREs with the LDA (thick-black lines) and SLDA (thin-red lines).}
\label{fig3}
\end{figure}

Finally, we investigate the differential conductance as a function of bias voltage at different temperatures, as shown in 
Fig.~\ref{fig3}. It is found that the unambiguous difference between the two approximations disappears with increasing bias 
voltage due to the decoherence effect introduced by nonequilibrium.

\section{Transient dynamics}

In this section, we perform numerical investigations on the spin-independent transient response of the interacting QD under 
various time-varying bias voltages, including steplike and harmonic modulations, by using a fourth-order Runge-Kutta scheme to 
solve the ensuing SOQREs. Therefore, we have to propagate every spin-independent second-order ADME, $\Omega_{\eta p,\eta'p'}(t)$, 
with time, which contains totally $8N_p^2$ unknown real quantities to propagate. The memory requirement of the proposed method 
scales with $N_M=8N_p^2+4N_p+1$ and the computational time requirement scales with $N_T\times N_M$ ($N_T$ is the number of time 
steps). Notice that for lower temperature, more Pad\'e pole pairs need to be considered in the Pad\'e spectrum decomposition of 
the Fermi function to ensure the numerical convergence. In the following calculations, we will hence focus our attention on the 
transient current only at an intermediate temperature, $T=0.1\Gamma$, for the purpose of avoiding huge computational burden. It 
deserves to point out that since the main tunneling mechanism is the cotunneling at this temperature as discussed above, our 
following simulations provide us a cheap and reasonable benchmark for understanding the strong correlated effects on electronic 
transient tunneling through nanoscale systems. Besides, it is indicated in Figs.~\ref{fig2} and \ref{fig3} that the SLDA is 
sufficiently accurate at this temperature. Then we will employ this approximation in the time evolution calculations.


\subsection{Response to steplike modulation}

In this subsection, we investigate the response of the interacting QD system to a large rectangular pulse bias potential of 
duration $s$. We assume that the chemical potentials of both leads are zero initially, and at $t=0$, a bias pulse with amplitude 
$\Delta_L$ is suddenly applied to the left lead, and increases the single-level energy of the QD by $\Delta_d$, but leaves the 
right lead unbiased. At a later time $t=s$ the pulse is switched off. This means following time dependences:
\bn
\epsilon_{d\sigma}(t) &=& \epsilon_d + [\theta(t)-\theta(t-s)]\Delta_d,\cr
\Delta_{L}(t) &=& [\theta(t)-\theta(t-s)]\Delta_{L},\cr
\Delta_{R}(t) &=& 0. \nonumber
\en
We further assume that the barrier heights do not depend on time, i.e., $u_{\eta}(t)\equiv 1$.

\begin{figure}[htb]
\includegraphics[height=6cm,width=8.5cm]{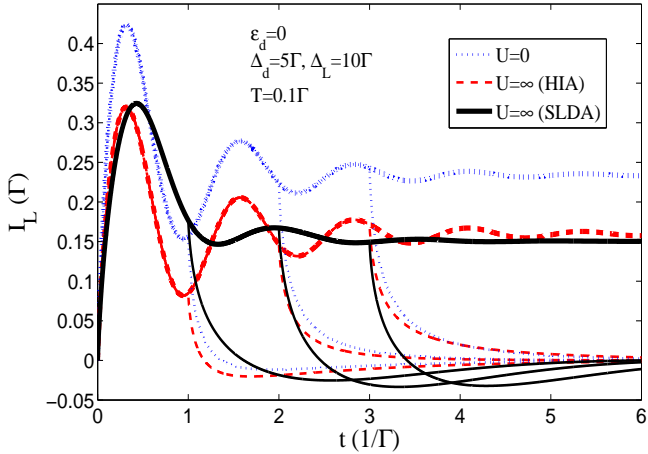}
\caption{(Colour online) Time-dependent current $I_L(t)$ through left lead in response to an upward step pulse with different 
durations $s=\infty$, $3/\Gamma$, $2/\Gamma$, and $1/\Gamma$, for the system with $\epsilon_d=0$ at the temperature $T=0.1\Gamma$ 
by the SOAREs with the SLDA (black-solid line) and the HIA (red-dashed line). In comparison, we also plot the results of a 
noninteracting single-level QD with the SOQREs (blue-dotted line).}
\label{fig4}
\end{figure}

In order to compare with the previous relevant work of exact EOM solutions, we first compute the current responses to step pulses 
with several durations for both the noninteracting single-level QD ($U=0$) and the interacting QD ($U=\infty$) with the same 
energy level $\epsilon_d=0$ at the temperature $T=0.1\Gamma$. The calculated time evolutions of current are plotted in 
Fig.~\ref{fig4} for the parameters: $\Delta_L=10\Gamma$ and $\Delta_{d}=5\Gamma$. It is clear that the simulations for the 
noninteracting system exactly recover the previous results of exact EOM solutions:\cite{Jauho,Maciejko} a ringing in the response 
current as a result of coherent electronic transitions between the leads and the QD with a pulse-amplitude-dependent 
period\cite{Jauho}
\bq
\Delta t_L= \frac{2\pi \hbar}{\mid \Delta_L - (\epsilon_d+\Delta_d) \mid}. \label{period}
\eq
For the interacting system the current with the HIA behaves similarly as the noninteracting system, exhibiting ringing with the 
same period $\Delta t_L=1.26/\Gamma$, but with an overall reduced amplitude. Nevertheless, the current response changes greatly 
if the correlation effect is considered more rigorously with the SLDA as shown by the black-solid line in Fig.~\ref{fig4}. It is 
found, such as, that the oscillation period is altered due to the level renormalization of the QD by the two-electron correlation 
tunneling as analyzed in the following discussions; the response time of the first peak is slower because the repulsive Coulomb 
interaction inhibits electron motion through the QD; the ringing fades away more quickly, which is a consequence of account of 
the bias-voltage-induced decoherence effect within the SLDA; and a temporary sign reversal of current is shown immediately after 
the pulse ends, which is absent in the HIA and noninteracting cases.

\begin{figure}[htb]
\includegraphics[height=6cm,width=8.5cm]{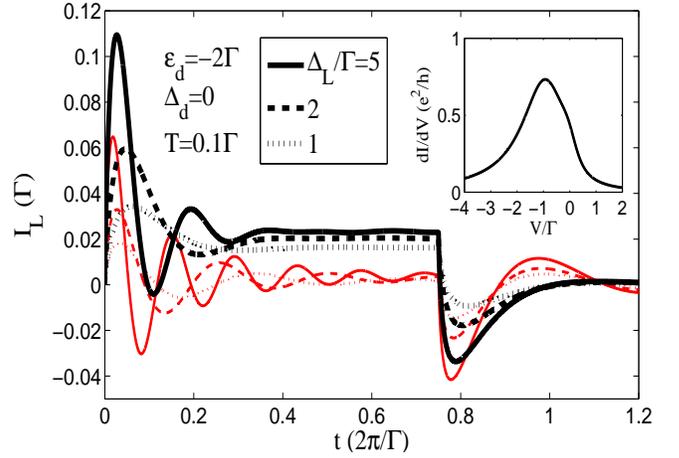}
\caption{(Colour online) Time-dependent current $I_L(t)$ through left lead in response to an upward step pulse of the duration 
$s=1.5\pi/\Gamma$
for the system with $\epsilon_d=-2\Gamma$ by the SLDA (black lines) and HIA (red lines). The current is driven by different 
applied bias voltages: $\Delta_L/\Gamma=5$ (solid lines), $2$ (dashed lines), and $1$ (dotted lines). The other parameters: 
$\Delta_d=0$, $T=0.1\Gamma$. Inset: The corresponding differential conductance calculated by the SOQREs with SLDA is plotted as a 
function of bias voltage in the case of steady-state transport.}
\label{fig5}
\end{figure}

In Fig.~\ref{fig5}, we study the current response behavior in the case that the energy of the QD level always remains at the 
Kondo regime, $\epsilon_d=-2\Gamma$, even as the bias pulse is applied (i.e. $\Delta_d=0$), to analyze further the effect of 
correlation on transient dynamics. It is clear that the HIA calculations predict the same period of current oscillation as 
indicated by Eq.~(\ref{period}), for example, $\Delta t_L=2\pi/7\Gamma$ for the case of $\Delta_L=5\Gamma$, nevertheless the SLDA 
calculations show that the current oscillates with a different period $\Delta t_L=2\pi/6\Gamma$. The SLDA calculation also 
disagrees with the time-dependent NCA calculation, which predicts that the ringing period is independent of dot level 
$\epsilon_d$, but instead is dependent only on the bias voltage, $2\pi/\Delta_L$.\cite{Plihal} This specific period has been 
ascribed to the Kondo peak splitting. The nonequilibrium DOS of the Kondo dot exhibits the splitting of the Kondo peak into two 
resonances, one at the Fermi energy of each lead. The coherent electronic transitions between the two peaks induce current 
oscillation with a period $2\pi \hbar/\mid \mu_L+\Delta_L-\mu_R-\Delta_R \mid = 2\pi /\Delta_L$.\cite{Plihal} However, since we 
consider the temperature $T=0.1\Gamma$ being much higher than the corresponding Kondo temperature, $T_K=0.0042\Gamma$, no Kondo 
peak appears in the equilibrium DOS of the dot electrons, as clearly shown in the inset of Fig.~\ref{fig5}. Interestingly, a wide 
peak is found nearly at $V=-1.0\Gamma$, which indicates that two-electron correlated tunneling causes an obvious renormalization, 
$\widetilde\epsilon_d\simeq -1.0\Gamma$, of the single-particle energy level of the dot. As far as this level renormalization is 
concerned, we can still ascribe our new ringing period to coherent electronic transitions between the leads and the QD with the 
renormalied level, that is
\bq
\Delta t_L= \frac{2\pi \hbar}{\mid \Delta_L - (\widetilde\epsilon_d+\Delta_d) \mid}.
\eq

\subsection{Response to harmonic modulation}

Now we move to investigate another time-dependent transport when external bias voltage is periodic in time. To be specific, we 
assume that the ac bias voltages is applied in-phase to the leads and dot as
\bq
\Delta_{d,\eta}(t)=\Delta_{d,\eta} \cos (\omega_{ac} t).
\eq

\begin{figure}[htb]
\includegraphics[height=6cm,width=8.5cm]{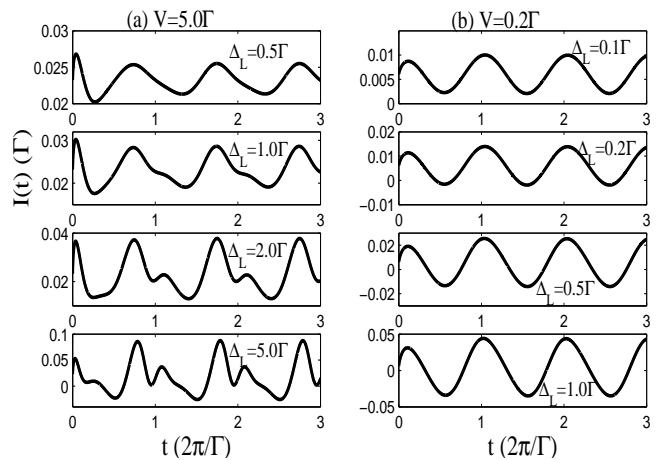}
\caption{(Colour online) The time evolutions of the current for harmonic modulations with different driving voltages $\Delta_L$ 
for two dc bias voltages, $V=5\Gamma$ (a) and $0.2\Gamma$ (b). The other parameters: $\epsilon_d=-2\Gamma$, $T=0.1\Gamma$, 
$\Delta_d=\Delta_R=0$. The modulation frequency is $\omega_{ac}=3\Gamma$.}
\label{fig6}
\end{figure}

\begin{figure}[htb]
\makebox{\includegraphics[height=2.92cm,width=4.5cm]{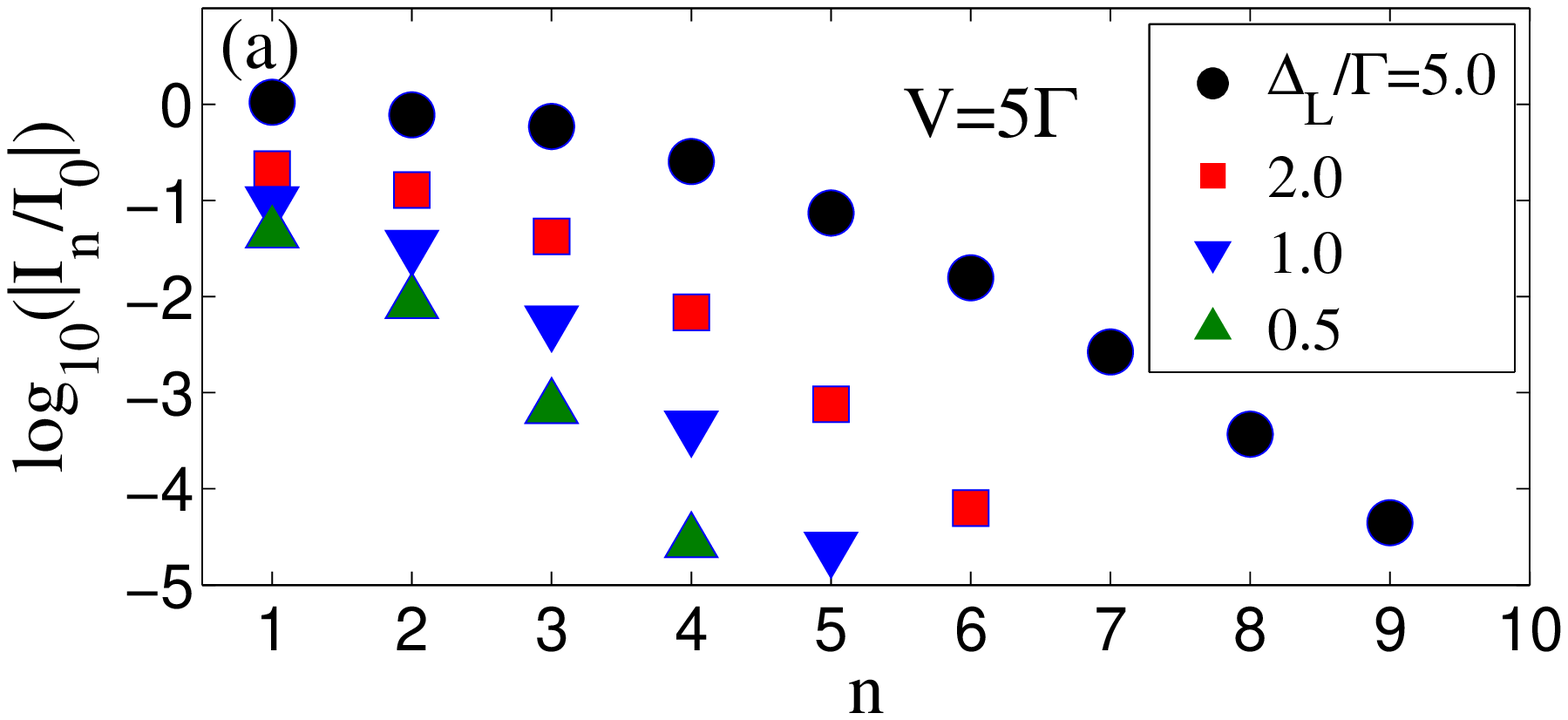}
         \includegraphics[height=3cm,width=4cm]{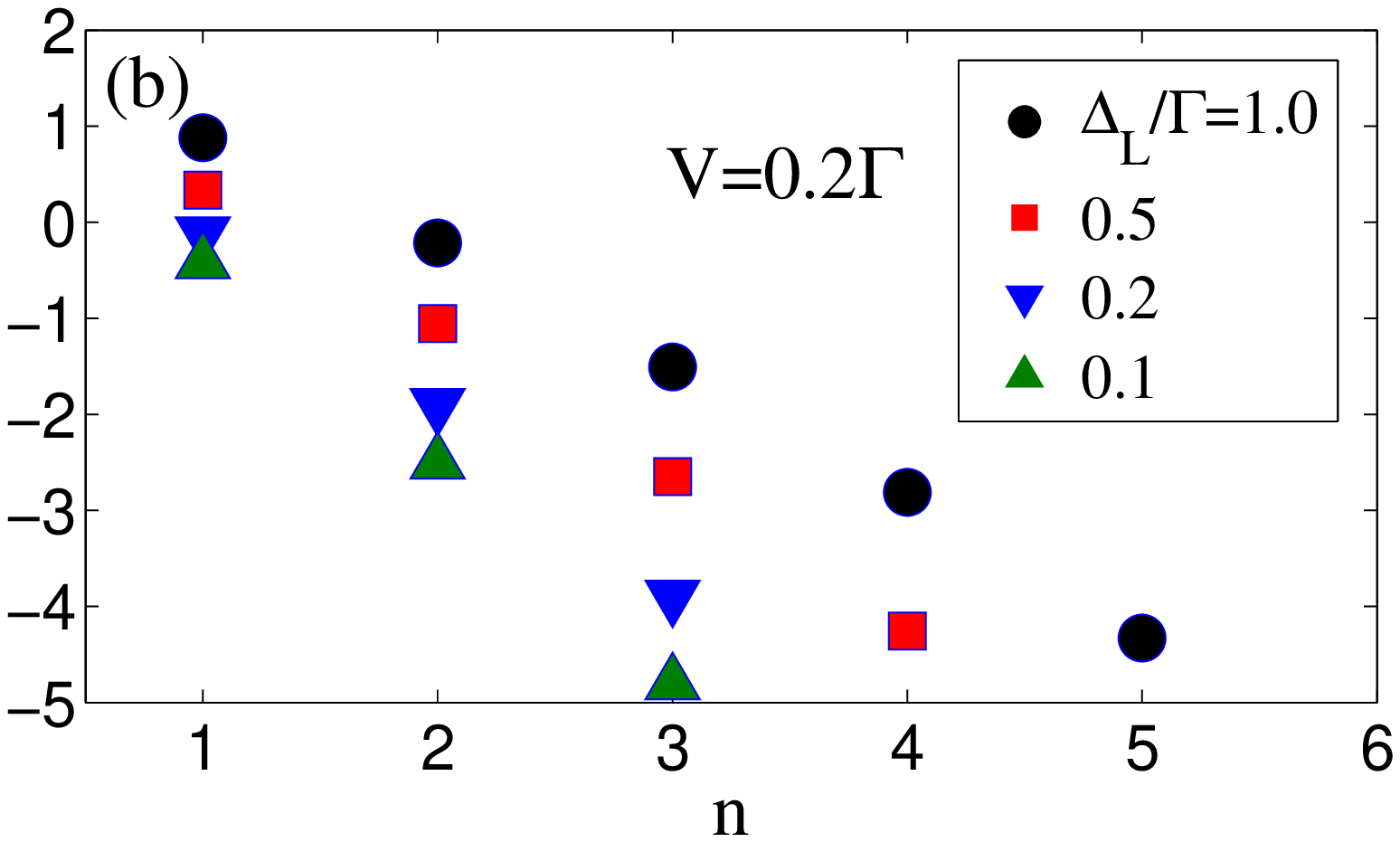} }
\caption{(Colour online) The Fourier representations of the current with different driving voltages $\Delta_L$ for two dc bias 
voltages, $V=5\Gamma$ (a) and $0.2\Gamma$ (b). The other parameters are the same as those in Fig.~\ref{fig6}.}
\label{fig7}
\end{figure}

Figure \ref{fig6} exhibits the time evolutions of current under harmonic bias voltages with various driving amplitudes $\Delta_L$ 
superposed to a dc bias voltage $V$ at the temperature $T=0.1\Gamma$. At the case of high dc bias voltage, $V=5.0\Gamma$, the 
response current shows the rich structure as displayed in Fig.~\ref{fig6}(a) that many harmonics are excited except for the case 
of low driving amplitude $\Delta_L=V/10=0.5\Gamma$. At the linear transport regime, the current response is different from that 
at the nonlinear regime. For instance, Fig.~\ref{fig6}(b) shows that only the dc and the first harmonic are enhanced in the 
current response at the case of low dc bias voltage, $V=0.2\Gamma$, even subject to a relatively strong driving voltage up to 
$\Delta_L=5V=1.0\Gamma$. This conclusion can be further confirmed by numerical calculation of the Fourier spectrum of the 
time-dependent current. We display the corresponding spectrum analysis (the ratio of amplitude of the $n$th harmonic component of 
the current, $I_n$, to the dc part, $I_0$) in Fig.~\ref{fig7} for the two cases. The current response to a harmonic modulation 
was previously studied for the Kondo model based on the Schrieffer-Wolff unitary transformation by means of bosonization method 
at the Toulouse limit and perturbation theory, respectively.\cite{Schiller1,Goldin} The two analytical solutions for the ac Kondo 
model predicted controversial results on the transient current spectrum. The present investigation provides current spectrum 
behavior that is different from both previous results. From Fig.~\ref{fig7}, we can state that at temperatures above the Kondo 
temperature, the magnitudes of the dc and ac components are governed by two independent parameters, the dc bias voltage $V$ and 
the driving voltage $\Delta_L$; the first harmonic component has a similar magnitude as the dc component and has at least one 
order of stronger magnitude than higher harmonics at the linear dc transport regime [Fig.~\ref{fig7}(b)]; while the second and 
even several higher harmonics are equally enhanced as the first harmonic component at the nonlinear dc transport regime 
[Fig.~\ref{fig7}(a)]. Therefore, our results show that the response current in the cotunneling regime has similar frequency 
spectrum behavior as the noninteracting system (of course the current is largely enhanced by the electron correlation effect).

\begin{figure}[t]
\includegraphics[height=6cm,width=8.5cm]{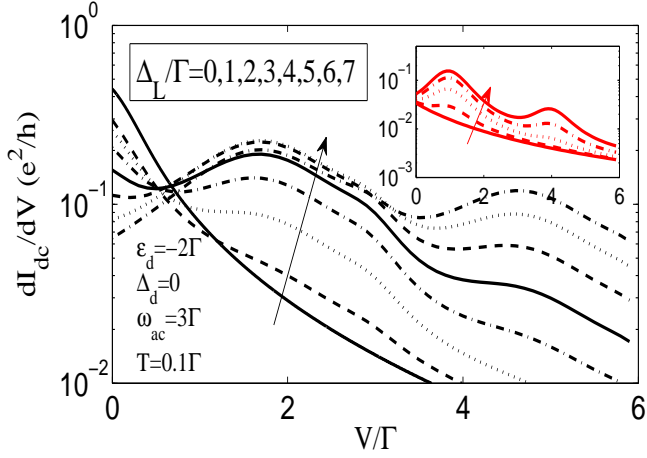}
\caption{(Colour online) The calculated dc differential conductance $dI_{dc}/dV$ versus dc bias voltage $V$ subject to a harmonic 
modulation with the frequency $\omega_{ac}=3\Gamma$ and various amplitudes $\Delta_L$ for the system with $\epsilon_d=-2\Gamma$ 
at the temperature $T=0.1\Gamma$. The other parameters: $\Delta_d=\Delta_R=0$. Inset: The corresponding results of the 
noninteracting QD.}
\label{fig8}
\end{figure}

In experiments, an easily measurable quantity is the dc differential conductance, which exhibits the famous zero-bias anomaly as 
one of the hallmarks of the nonlinear Kondo effect and electron correlated tunneling (Fig.~\ref{fig3}). In the presence of strong 
ac modulation, the dc differential conductance, defined as the derivative of time-average current $I_{dc}$ with respect to dc 
bias voltage $V$, will exhibit side peaks with increasing dc bias voltage that is ascribed to the photon absorption assisted 
tunneling, as shown in Fig.~\ref{fig8}, in which we plot the dc $dI_{dc}/dV$ versus dc bias voltage at different amplitudes of 
the driving field. At a sufficiently large modulation frequency, the noninteracting QD has resonances at $V=\epsilon_d+n\hbar 
\omega_{ac}$ (see the inset of Fig.~\ref{fig8}), while the previous analytical solutions for the Kondo model suggested different 
characteristic resonances located at $V=n\hbar \omega_{ac}$ resulted from the splitting of the Kondo peak in the nonequilibrium 
situation.\cite{Schiller1,Goldin} However, we can see clearly from Fig.~\ref{fig8} that at the temperature where the cotunneling 
dominates, the resonances occur at different locations, $V=\widetilde\epsilon_d + n\hbar \omega_{ac}$, due to the electron 
correlation induced level renormalization as indicated in the above subsection, if the driving field is sufficiently strong. 
Besides, in contrast to the noninteracting system, the zero-bias anomaly is suppressed by the harmonic modulation, even leading 
to a zero-bias minimum.\cite{Goldin,Kaminski,Lopez1}

\subsection{Linear response}

Another important issue in the context of the time-dependent transport in nanoscale devices is the current response to a small ac 
voltage driving, which is applied across the central region in addition to a dc bias voltage, i.e., the linear ac conductance in 
a nonequilibrium situation for particularly the interacting systems. The present SOQREs facilitate the calculation of the current 
response in the limiting case of a slight ac amplitude, $\Delta_{L}(t)=-\Delta_{R}(t)=\frac{1}{2}\delta V \cos(\omega t)$ with 
$\delta V\rightarrow 0$.

For the sake of convenience, we rewrite the SOQREs with SLDA in matrix form:
\bq
\frac{\partial {\bm \varrho}}{\partial t} = [{\bm A}_0 + {\bm A}_1 \delta V \cos(\omega t)] {\bm \varrho} + {\bm B}, 
\label{eommatrix}
\eq
where ${\bm \varrho}=(\rho_{\sigma\sigma}, \Pi_{\eta \sigma p}, \Pi_{\eta\sigma p}^\dagger, \Omega_{\eta \sigma p, \eta' 
\bar\sigma p'}, \Omega_{\eta\sigma p, \eta'\bar\sigma p'}^\dagger)^{\rm T}$ and the matrices ${\bm A}_0$, ${\bm A}_1$, and ${\bm 
B}$ can be read from Eqs.~(\ref{dotrho}), (\ref{eomcmame4}), and (\ref{eomsoadmesla}). Under the condition of small oscillation 
amplitude, the time-dependent ADMEs ${\bm \varrho}$ can be expanded as\cite{Dong2}
\bq
{\bm \varrho}= {\bm \varrho}^{(0)} + ({\bm \varrho}^{(11)} e^{i\omega t} + {\bm \varrho}^{(1\bar 1)} e^{-i\omega t} )\delta V ,
\eq
where ${\bm \varrho}^{(0)}$ represents the zeroth order solution irrespective of the ac field, i.e., the stationary solution 
without ac modulation, and ${\bm \varrho}^{(1\pm 1)}$ is the positive (negative) frequency part of the first order correction to 
the stationary solution. Substituting this expansion into the time-dependent SOQREs Eq.~(\ref{eommatrix}) and expanding according 
to the perturbation parameter $\delta V$, therefore, we obtain\cite{Dong2}
\bn
{\bm \varrho}^{(0)} &=& -{\bm A}_0^{-1} {\bm B} ,\\
{\bm \varrho}^{(1\pm 1)} &=& (\pm i\omega {\bm I} - {\bm A}_0)^{-1} {\bm A}_1 {\bm \varrho}^{(0)},
\en
with ${\bm I}$ being the unit matrix. Using the first order correction term ${\bm \varrho}^{(1\pm 1)}$ we can calculate the 
linear response current as
\bn
\delta I_{\eta \sigma}(t) &=& \left \{ -\frac{3}{2} \Gamma_{\eta \sigma} \rho_{\sigma \sigma}^{11}+ \sum_{p} \left [ \Pi_{\eta 
\sigma\sigma p}^{11} + \left ( \Pi_{\eta \sigma p}^{1\bar 1} \right )^* \right ] \right \} e^{i\omega t} \delta V \cr
&& \hspace{-1.5cm} + \left \{ -\frac{3}{2} \Gamma_{\eta \sigma} \rho_{\sigma \sigma}^{1\bar 1}+ \sum_{p} \left [ \Pi_{\eta 
\sigma\sigma p}^{1\bar 1} + \left ( \Pi_{\eta \sigma p}^{11} \right )^* \right ] \right \} e^{-i\omega t} \delta V.
\en
Then we perform Fourier transformation of the total current $\delta I(t)= \sum_{\sigma}(\delta I_{L\sigma}+ \delta 
I_{R\sigma})/2$ and obtain the linear response admittance $G(\omega)$ of the QD
\bq
G(\omega)= \frac{\delta I(\omega)}{\delta V}.
\eq

\begin{figure}[htb]
\includegraphics[height=5.5cm,width=8.5cm]{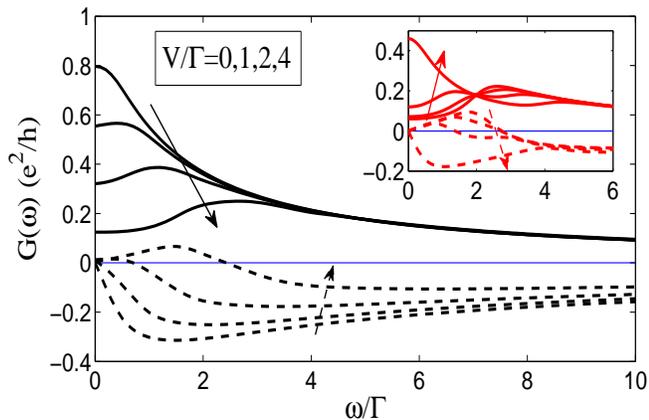}
\caption{(Colour online) Real $\Re G(\omega)$ (solid lines) and imaginary $\Im G(\omega)$ (dashed lines) parts of the linear 
response admittance, $G(\omega)$, versus frequency for different dc bias voltages $V$ for the QD with $\epsilon_d=-2\Gamma$ at 
the temperature $T=0.1\Gamma$. Inset: The corresponding results for the noninteracting system.}
\label{fig9}
\end{figure}

In Fig.~\ref{fig9} we show the resulting linear response admittance $G(\omega)$ at the linear transport ($V=0$) and nonlinear 
transport (a finite dc bias voltage $V\neq 0$ is applied symmetrically between the left and right leads, i.e., 
$\mu_L=-\mu_R=V/2$) regimes. Note that our sign of the imaginary part of the admittance is opposite to the sign in 
Ref.~\onlinecite{Fu} and that a positive (negative) $\Im G(\omega)$ corresponds to the sign of a capacitive (inductive) 
component. For zero external dc bias voltage, the calculated admittance for the noninteracting system resembles exactly the 
results reported by Fu and Dudley:\cite{Fu} For the QD with $\epsilon_d=-2\Gamma$, the admittance is capacitive at low 
frequencies, but turns to an inductive behavior at $\omega\sim \mid \epsilon_d \mid$, meanwhile the ac conductance $g(\omega)=\Re 
G(\omega)$ exhibits a peak implying that a photon-assisted resonance is satisfied at this condition. Moreover, present 
investigation shows that a finite dc bias voltage will change the crossover point as $\omega\sim \mid \epsilon_d - \mu_R \mid = 
\mid \epsilon_d + V/2 \mid$ since a new photon-assisted resonance is formed between the level and the right lead. At the case of 
$V=4\Gamma$, the admittance becomes always inductive behavior and a monotonically decreasing of the ac conductance. It is 
nevertheless observed that the bias voltage dependence of the admittance exhibits contrary behavior for the interacting QD in the 
cotunneling regime. The external bias voltage induces suppression of the ac conductance at low frequencies. The equilibrium 
admittance is always inductive but becomes capacitive at low frequencies due to external dc bias driving. Therefore, we can 
conclude that the correlation effect greatly changes the linear ac-response behavior of the QD.

\section{Conclusion}

In this paper, we have presented a convenient approach for the time-dependent electronic Kondo-type tunneling and cotunneing 
through an interacting QD subject to arbitrary time-varying potentials based on the TDNGF. For this purpose, we have followed the 
theoretical strategy of a recently developed propagation scheme for time-dependent transport in a noninteracting QD, in which a 
set of coupled ordinary differential equation with only one time argument is derived based on the double-time TDNGF by the help 
of an auxiliary-mode expansion.\cite{Croy1} Unlike the case of noninteracting QD, the second-order ADMEs of the interacting 
system are however found to be related to the two-particle GFs, whose EOMs will inevitably generate the next-order GFs due to the 
on-site Coulomb interaction. Then we have employed the famous LDA and it's simplified version to evaluate the higher order GFs. 
At the end, a closed set of coupled single-time evolution equations called SOQREs has been established.
It deserves to notice that, in comparison with the previous SOvN approach, the SOQREs can provide correct description for 
transient dynamics of the correlated electron transport, since no Markov approximation is invoked in its derivation. Meanwhile, 
the SOQREs deal with the electron correlation more rigorously because this scheme is closely joint with the NGF technique, which 
is believed to be most powerful method to investigate the strongly electronic correlation effect.

To validate the present approach, we have first utilized the resulting differential equations to investigate the steady-state 
transport. Our numerical calculations show that the SOQREs with the original LDA is valid at the whole range of temperature, from 
the Kondo-type tunneling regime to the sequential tunneling regime, while the SOQREs with the SLDA is only valid at temperatures 
above the Kondo temperature since this simplified decoupling approximation takes no account of the back action effect of electron 
tunneling on the electrodes.

As applications, we have employed the present SOQREs with the SLDA to investigate time-dependent cotunneling through an 
interacting QD in response to abrupt change and a harmonic modulation of bias voltage, respectively. Our calculations have 
revealed some new interesting physics in the current response behavior: 1) a ringing of current with a new period after a sharp 
bias turn-on; 2) similar frequency spectrum response of the ac-driving current with the noninteracting QD; but 3) different 
time-average current-dc bias voltage characteristic from the noninteracting QD, such as suppression of zero-bias peak in the dc 
differential conductance with increasing harmonic modulation amplitude and a new photon-assisted resonant condition. Our further 
analysis has indicated that these features are ascribed to an effective renormalization of the dot level due to the interplay of 
electron correlation and coherent tunneling between the QD and leads. Finally, we have discussed ac linear-response at arbitrary 
dc bias voltage and predicted that the interacting QD exhibits entirely different linear-response admittance from the 
noninteracting system.

Before ending this paper, let us mention a few possible directions of future research. First, a WBL has been invoked in the 
derivation of the SOQREs in the present work. Such model has a great advantage that it is very easy for theoretical calculation 
and can still provide some insight into the physics of time-dependent transport through a QD, even though it is far from a 
realistic device. The present formulation, nevertheless, can be readily extended to the case of a finite bandwidth, such as the 
Lorentzian linewidth. Moreover, it is known that the electronic spectral density inside the QD satisfies the Fermi liquid unitary 
limit $\varrho_d(\omega=0)=2/\Gamma$ (This DOS implies that the linear conductance reaches its unitary limit, $2e^2/h$.) when the 
temperature goes to
zero.\cite{Entin,Kashcheyevs,Roermund} However, the present approach can not reach such unitary limit since the LDA results in a 
logarithmic-$T$ dependence of the DOS in the low-$T$ limit but not the exact $T^2$ dependence.\cite{Entin,Kashcheyevs} Therefore, 
the third-order quantum rate equations approach based on a more rigorous decoupling approximation beyond the LDA,\cite{Roermund} 
is desirable for future investigation of time-dependent Kondo physics at extremely low temperature.

\begin{acknowledgments}

This work was supported by Projects of the National Basic Research Program of China (973 Program) under Grant No. 2011CB925603, 
and the National Science Foundation of China, Specialized Research Fund for the Doctoral Program of Higher Education (SRFDP) of 
China.

\end{acknowledgments}

\appendix*
\section{Pad\'e expansion of the Fermi function}

In order to perform the energy integration in the self-energies Eqs.~(\ref{selfen1}) and (\ref{selfen2}), it is needed to expand 
the Fermi function in terms of a finite sum over simple poles. The Pad\'e spectrum decomposition is believed to be most efficient 
in numerical calculation, which is defined as follows:\cite{Karrasch,Hu,Xie}
\bn
f(x) &=& \frac{1}{1+e^x} = \frac{1}{2} - \frac{1}{2} \frac{\sinh(x/2)}{\cosh(x/2)} \cr
&\approx & \frac{1}{2} - \sum_{p=1}^{N_p} R_p \left ( \frac{1}{x-i\chi_p} + \frac{1}{x+i\chi_p} \right ),
\en
where $R_p$ is the $p$th residue in the Pad\'e expansion, $i\chi_p$ ($\chi_p>0$) is the $p$th Pad\'e pole in the upper complex 
plane, and $N_p$ is the number of Pad\'e pole pairs. For this decomposition, the residues and poles can be given by the 
eigenvalue problem of the symmetric matrix $B$,\cite{Karrasch}
\bq
B\mid b_p\rangle = b_p \mid b_p \rangle,
\eq
with
\bq
B_{n,n+1} = \frac{1}{2\sqrt{(2n-1)(2n+1)}}, \quad n\geq 1,
\eq
as
\bn
\chi_p &=& 1/b_p, \\
R_p &=& \mid \langle 1 \mid b_p \rangle \mid^2/(4b_p^2).
\en

\end{document}